\documentclass[final]{siamart}

\usepackage[latin1]{inputenc}
\usepackage{graphicx,amssymb,mathrsfs,amsmath,amsfonts,appendix}
\usepackage{booktabs}
\usepackage{color}
\usepackage{algorithm,algorithmic,nccmath,enumerate,psfrag,forloop,ifthen,soul}
\usepackage{stackrel}
\usepackage{multirow}
\usepackage{hyperref}
\usepackage{dsfont}
\usepackage{todonotes}
\usepackage{enumitem}

\newcommand{\bfx}{{\bf x}}
\newcommand{\ud}{\mathrm{d}}	
\newcommand{\bo}[1]{\boldsymbol{#1}} 

\newcommand{\TheTitle}{Diffusion of particles with short-range interactions} 
\newcommand{\TheAuthors}{M. Bruna, S. J. Chapman, and M. Robinson}

\headers{\TheTitle}{\TheAuthors}

\title{{\TheTitle}\thanks{This work was partly funded by an EPSRC Cross-Discipline Interface Programme (grant EP/I017909/1), St John's College Research Centre and the John Fell Fund.}}

\author{
  Maria Bruna\thanks{Mathematical Institute, University of Oxford, Radcliffe Observatory Quarter, Woodstock Road,
Oxford OX2 6GG, United Kingdom
    (\email{bruna@maths.ox.ac.uk}, \email{chapman@maths.ox.ac.uk}).}
  \and
  S. Jonathan Chapman\footnotemark[2]
  \and
  Martin Robinson\thanks{Department of Computer Science, University of Oxford, Parks Road, Oxford OX1 3QD, United Kingdom (\email{Martin.Robinson@cs.ox.ac.uk}).}
}

\usepackage{amsopn}

\ifpdf
\hypersetup{
  pdftitle={\TheTitle},
  pdfauthor={\TheAuthors}
}
\fi

\begin{document}

\maketitle

\begin{abstract}
A system of
interacting Brownian particles subject to short-range repulsive
potentials is considered. A continuum description in the form of a
nonlinear diffusion equation is derived systematically in the dilute
limit using the method of matched asymptotic expansions.  Numerical simulations
are performed to compare the results of the model with those of the
commonly used   mean-field and Kirkwood-superposition approximations,
as well as with Monte Carlo simulation of the stochastic particle
system, for various interaction potentials.
 Our approach works best for very repulsive
short-range potentials, while the mean-field approximation  is suitable for
long-range interactions. The Kirkwood superposition approximation provides
an accurate description for both short- and long-range potentials, but
is considerably more computationally intensive.
\end{abstract}



\section{Introduction}

Nonlinear diffusion equations are often used to describe a system of
interacting particles at the continuum level. These play a key role in
various physical and biological applications, including colloidal
systems and granular gases, ion transport, chemotaxis, neural
networks, and animal swarms. These continuum models are important as
tools to explain how individual-level mechanisms give rise to
population-level or collective behavior.  
Closure approximations such as the mean-field closure are often used to obtain the continuum model, but depending on the type of interactions they can lead to substantial errors. In this paper we present a new approach that is suited for short-range repulsive interactions. 

A typical model for a system of $N$ interacting particles is to assume
each particle evolves according to the overdamped Langevin dynamics
and interacts with the other particles via a pairwise interaction
potential $u$, so that
\begin{equation} 
\label{ssde}
\ud {\bf X}_i(t) =  \sqrt{2D} \, \ud{\bf W}_i(t) + {\bf f}  ( {\bf X}_i (t) ) \ud t  - \sum_{j \ne i}  \nabla_{{\bf x}_i} u(\|{\bf X}_i(t) - {\bf X}_j(t)\|) \ud t, 
\end{equation}
for $i = 1, \dots, N$,
where ${\bf X}_i (t) \in \Omega \subset \mathbb R^d$ is the position of the $i$th particle at time $t$,  $D$ is the diffusion constant, ${\bf W}_i(t)$ denotes a $d$-dimensional Brownian motion, and ${\bf f}: \Omega \to \mathbb R^d$ is an external force. 

Despite the conceptual simplicity of the stochastic model \cref{ssde}, it can be computationally intractable for systems with a large number of interacting particles $N$, since the interaction term has to be evaluated for all particle pairs. In such cases, a continuum description of the system, based on the evolution of the population-averaged spatial concentration instead of individual particles, becomes attractive. Depending on the nature of the interaction potential different averaging techniques may be suitable. Interactions can be broadly classified into local and nonlocal depending on the range of interaction between particles. Nonlocal interactions are associated with a long-range or \emph{ultra-soft} interaction potential $u$  in \cref{ssde}. Then every particle can interact not only with its immediate neighbors but also with particles far away, and one can use a mean-field approximation to obtain a partial differential equation (PDE) for the one-particle probability density $p({\bf x},t)$ of finding a given particle at position $\bf x$ at time $t$. The standard mean-field approximation (MFA)  procedure applied to the microscopic model \cref{ssde} gives
\begin{equation} \label{mean_field}
\frac{\partial p}{\partial t}  =  \nabla_{\bf x} \left [ D  \nabla_{\bf x} p - {\bf f} ({\bf x}) p  + (N-1) p \nabla_{\bf x} (u\ast p) \right],
\end{equation} 
where  $u \ast p = \int u(\|{\bf x} - {\bf y}\|) p({\bf y}) \ud {\bf
  y}$. The main assumption in writing down \cref{mean_field} is that,
in deriving the interaction term,  particles can be treated as though
they were uncorrelated. Since one is often interested in system with a
large number of particles $N$, it is common to consider the number
density ($\rho = N p$) and take the limit in which the number of
particles and the volume tend to infinity, while keeping the average
number density constant, that 
is, $N, V\to \infty$, $N/V = \rho_0$. This is known as the thermodynamic
limit \cite[\textsection 2.3]{Hansen:2006uv}, and results in equation
(\ref{mean_field}) for $\rho$ without the factor $N-1$.

While the mean-field approximation \cref{mean_field} is convenient and leads to an accurate description for long-range interactions, such as with Coulomb interactions \cite{Dolbeault:2001jc}, it fails when considering relatively strong repulsive short-range potentials. Sometimes \cref{mean_field} does not make sense since the convolution does not exist; in other cases, it results in a poor model of the system because the underlying assumptions of the method are not satisfied, as we will discuss later.  In particular, the model \cref{mean_field} does not make sense with hard-core repulsive interactions, which are commonly used to model excluded-volume effects in biological and social contexts \cite{Bruna:2012cg}. A common way to circumvent this is to assume particles are restricted to a lattice, giving rise to so-called on-lattice models. The most common of these is the simple exclusion model, in which a particle can only move to a site if it is presently unoccupied \cite{liggett1999stochastic}. One can derive an analogous continuous limit to \cref{mean_field} using Taylor expansions \cite{Burger:2010gb}. However, it turns out that with identical particles the interaction terms (analogous to the convolution term in \cref{mean_field}) cancel out unless the external force $\bf f$ is nonzero. 

Because of the issues that the MFA has in particle systems with short-range interaction potentials, one must often resort to numerical regularisations \cite{Horng:2012io} or alternative closure approximations. There are a large number of closure approximations, and choosing the right closure for a given pair potential is ``an art in itself'' \cite{Singer:2004ec} due to the phenomenological nature of the approach. Each choice of closure relation results in a different approximate equation for the density $p$, and it is not clear a priori whether it will be a good approximation. As discussed above, one may even obtain an equation that does not make sense or is ill-posed. 

One class of closures, including the MFA and the Kirkwood superposition  approximation (KSA), impose a relation between the $n$th and $(n+1)$th density functions in the BBGKY hierarchy (see \cref{sec:IBM}). Whereas the MFA closes at $n=1$ (writing the two-particle density function in terms of the one-particle density), KSA closes at $n=2$, approximating the three-particle density function as a combination of the one- and two-particle density functions \cite{Kirkwood:1935is}. While the KSA originated in the field of statistical mechanics, where it has been the basis of a whole theory, it has recently also been used in biological applications to obtain closed equations for a system of biological cells \cite{Binny:2015ei,Middleton:2014fa}. However, the resulting KSA model is quite complicated to solve and the MFA remains the most commonly used approximation. 

Another class of closure relations, similar in spirit to KSA, is based on the Ornstein--Zernike (OZ) integral equation \cite{Hansen:2006uv}. Here the pair correlation function is decomposed into a `direct' part and an `indirect' part. The latter is mediated through (and integrated over) a third particle. In addition, the OZ equation requires a further closure assumption providing an additional relation between the direct and indirect correlations. Commonly used closures, for hard spheres and soft spheres respectively, include the Percus--Yevick approximation and the hypernetted chain approximation. 

In this paper we are interested in systems such as \cref{ssde} with short-range repulsive interactions for which MFA fails. We will employ an alternative averaging method to obtain a continuum description of the system based on matched asymptotic expansions (MAE). Unlike the mean-field approach, this method is a systematic asymptotic expansion which does not rely on the system size being large. It is valid for low concentrations, exploiting a small parameter $\epsilon$ arising from the short-range potential and the typical separation between particles. The result is a nonlinear advection-diffusion equation of the form 
\begin{align}
\label{mae_eq}
\frac{\partial p}{\partial t} = \nabla_{
 \bf  x} \cdot   \left [ D \nabla_{\bf  x}  p -  {\bf f}({\bf  x}) p + \alpha_u \epsilon^d (N-1)  p \nabla_{\bf  x}  p   \right ],
\end{align}
where the coefficient $\alpha_u$ depends on the interaction potential $u$ and $d$ is the dimension of the physical space.

The remainder of the paper is organized as follows. In \cref{sec:IBM} we introduce the Fokker--Planck PDE for the joint probability density of the particle system; this is another individual-based description equivalent to the Langevin stochastic differential equation \cref{ssde}. In \cref{sec:closure} we discuss three common closure approximations which reduce the Fokker--Planck equation to  a population-level PDE.  In \cref{sec:soft_mae} we present our alternative approach to closure based on MAE, and derive equation \cref{mae_eq}. In \cref{sec:results} we test the models obtained from the different methods against each other and stochastic simulations of the stochastic particle system  for various interaction potentials. Finally, in \cref{sec:conclusions}, we present our conclusions.

\section{Individual-based model} \label{sec:IBM}

We consider a set of $N$ identical particles evolving according to the Langevin stochastic differential equation \cref{ssde} in a domain $\Omega \subset \mathbb R^d$, with $d \le 3$. We nondimensionalise time and space such that the diffusion coefficient $D = 1$, and the volume of the domain  $| \Omega | = 1$. We suppose the interaction potential $u(r)$ is repulsive and short range, with range $\epsilon \ll 1$. 
The interaction potential of a system of $N$ particles is, assuming pairwise additivity, the sum of isolated pair interactions
\begin{equation} 
\label{interacpot}
U(\vec x) =  \sum_{1\le i<j\le N} u (\| {\bf x}_i - {\bf x}_j\|),
\end{equation}
where $\vec x = ({\bf x}_1, \dots, {\bf x}_N)$ is the $N$-particle position vector. The interaction force acting on the $i$th particle due to the other $N-1$ particles is given by
\begin{equation} 
\label{softinter}
{\bf g}_i (\vec x) = - \nabla_{{\bf x}_i} U (\vec x) = - \sum_{j \ne i} \nabla_{{\bf x}_i} u(\|{\bf x}_i - {\bf x}_j\|). 
\end{equation}
Here forces are non-dimensionalized with the mobility (the inverse of the drag coefficient) so that we can talk about a force acting on a Brownian particle.  Finally, we suppose that the initial positions ${\bf X}_i(0)$ are random and identically distributed.

The counterpart of \cref{ssde} in probability space is the Fokker--Planck equation
\begin{subequations}
\label{sfp}
\begin{align}
\label{sfp_eq}
\frac{\partial P}{\partial t} (\vec x, t) = \vec \nabla_{\vec x} \cdot \left [  \vec \nabla_{\vec x} P - \vec F (\vec x) P +  \nabla_{ \vec x} U (\vec x) P \right ] \qquad \text{in} \qquad \Omega^N,
\end{align}
where $P(\vec x, t)$ is the joint probability density function of the $N$ particles being at positions $\vec x = ({\bf x}_1, \dots, {\bf x}_N)\in \Omega^N$ at time $t$ and $\vec F (\vec x) = \bo({\bf f}_1({\bf x}_1), \dots, {\bf f}_N({\bf x}_N) \bo)$. Since we want to conserve the number of particles, on the domain boundaries $\partial \Omega ^N$ we require either no-flux or periodic boundary conditions. Throughout this work we use the latter. Accordingly, the potential $u$ will be a periodic function in $\Omega$. 
The initial condition is 
\begin{equation} \label{ini_P}
P(\vec x, 0) = P_0(\vec x),
\end{equation}
with $P_0$ invariant to permutations of the particle labels. 
\end{subequations}

We proceed to reduce the dimensionality of the problem \cref{sfp} by looking at the marginal density function of one particle (the first particle, say) given by  
\begin{equation}
p({\bf x}_1,t) = \int_{\Omega^{N-1}} P(\vec x, t) \, \ud {\bf x}_2 \cdots {\bf x}_N.
\end{equation}
The particle choice is unimportant since $P$ is invariant with respect to permutations of particle labels. Integrating \cref{sfp_eq} over ${\bf x}_2, \dots, {\bf x}_N$ and applying the divergence theorem gives
\begin{equation} 
\label{sfp_int}
\frac{\partial p}{\partial t} ({\bf x}_1,t) = \nabla_{{\bf  x}_1}  \cdot \left[ \nabla_{{\bf  x}_1} \,  p  - {\bf f}({\bf x}_1) p  +  {\bf G}({\bf x}_1,t)  \right],
 \end{equation}
where ${\bf G}$ is the $d$-vectorial function
\begin{align} 
\label{Balltrue}
\begin{aligned}
{\bf G}({\bf x}_1,t) &= \int_{\Omega^{N-1}}  P( {\bf x}_1, {\bf x}_2, \dots,  {\bf x}_N, t) \sum_{j=2}^N \nabla_{{\bf x}_1} u (\|{\bf x}_1 - {\bf x}_j\|)   \,  \ud {\bf x}_2 \cdots \ud {\bf x}_N \\
&= (N-1) \int_{\Omega} P_2( {\bf x}_1, {\bf x}_2, t) \nabla_{{\bf x}_1} u(\|{\bf x}_1 - {\bf x}_2 \|)   \,  \ud {\bf x}_2,
\end{aligned}
\end{align}
and
\begin{equation} 
P_2( {\bf x}_1, {\bf x}_2, t) = \int_{\Omega^{N-2}} P(\vec x, t) \,  \ud {\bf x}_3 \cdots \ud {\bf x}_N
\end{equation}
is the two-particle density function, which gives the joint probability density of particle 1 being at position ${\bf x}_1$ and particle 2 being at ${\bf x}_2$. An equation for $P_2$ can be written from \cref{sfp_eq}, but this then depends on $P_3$, the three-particle density function. This results in a hierarchy (the BBGKY hierarchy) of $N$ equations for the set of $n$-particle density functions ($n = 1, \dots, N$), the last of which is \cref{sfp_eq} itself (since $P$ is the $N$-particle density function). 
In order to obtain a practical model, a common approach is to truncate this hierarchy at a certain level to obtain a closed system. In particular, closure approximations in which the $n$-particle density function $P_n$ is replaced by an expression involving lower density functions $P_s$, $s<n$, are commonly used. However, because of their phenomenological nature, they can often lead to errors in the resulting model. In the next section we present three such closure approximations and highlight the issues they encounter when dealing with short-range repulsive potentials. In \cref{sec:soft_mae} we present an alternative approach based on matched asymptotic expansions.

\section{Closure approximations}
\label{sec:closure}

\subsection{Mean-field closure} \label{sec:easy_closure}

The simplest and most common closure approximation is to assume that particles are not correlated at all in evaluating the interaction term $\bf G$, that is,
\begin{align} 
\label{closure_approx}
P_2( {\bf x}_1, {\bf x}_2, t) = p( {\bf x}_1, t) p( {\bf x}_2, t).
\end{align}
Substituting \cref{closure_approx} into \cref{Balltrue} gives
\begin{align} 
\label{Bcclosed}
\begin{aligned}
{\bf G} ({\bf x}_1,t)= (N-1 ) \, p( {\bf x}_1,t) \! \int_{\Omega} p( {\bf x}_2, t) \nabla_{{\bf x}_1} u(\|{\bf x}_1 - {\bf x}_2\|)   \,  \ud {\bf x}_2.
\end{aligned}
\end{align}
Combining this with the equation for $p$ in \cref{sfp_int} gives equation \cref{mean_field} presented in the introduction, the mean-field approximation (MFA). However, one should keep in mind that \cref{closure_approx} might not always be valid when using such model. In particular, when $u(r)$ is a short-range interaction potential, the dominant contribution to the integral \cref{Balltrue} is when ${\bf x}_1$ is close to ${\bf x}_2$, and this is exactly the region in which the positions of particles are correlated. 
We note that the mean-field closure is often used implicitly with \cref{mean_field} written down directly rather than being derived from \cref{sfp_eq} \cite{Horng:2012io,Mogilner:1999iy}. The reasoning goes as follows: if $p({\bf x},t)$ is the probability of finding a particle at ${\bf x}$, the force on a particle at ${\bf x}_1$ is given by multiplying the force due to another particle at ${\bf x}_2$ by the density of particles at ${\bf x}_2$ and integrating over all positions ${\bf x}_2$. 

If we suppose the pair potential is short-ranged, we can approximate the integral in \cref{Bcclosed} to remove of the convolution term. In particular, we suppose that $u = O(r^{-(d+\delta)})$ for some $\delta>0$ as $r\to \infty$ and rewrite the potential as $u(r) = \tilde u(r/\epsilon)$ with $\epsilon \ll 1$.  Introducing the change of variable ${\bf x}_2 = {\bf x}_1 + \epsilon \tilde {\bf x}$ and expanding $p({\bf x}_2,t)$ about ${\bf x}_1$ gives
\begin{align} 
{\bf G} ( {\bf x}_1,t) = - \epsilon^{d-1} (N-1 ) \, p(  {\bf x}_1,t) \! \int_{\mathbb R^d}   \left[ p(  {\bf x}_1, t) + \epsilon  \tilde {\bf x} \cdot \nabla_{ {\bf x}_1} p(  {\bf x}_1, t)  \right]   \nabla_{\tilde {\bf x}} \,  \tilde u( \| \tilde {\bf x} \| ) \,   \ud \tilde {\bf x} + \cdots,
\end{align}
where we can extend the integral with respect to variable $\tilde {\bf x}$ to the whole space since the potential $\tilde u$ is localized near the origin and decays at infinity. 
Noting that the potential is a radial function, the leading-order term in the integral vanishes, and, after integrating by parts in the next term, we obtain
\begin{align} 
\label{Bcfinal}
{\bf G} ( {\bf x}_1,t) \sim \epsilon^{d} (N-1 ) \, p(  {\bf x}_1,t) \nabla_{ \tilde {\bf x}_1} p(  {\bf x}_1,t)   \int_{\mathbb R^d}  \tilde u( \| \tilde {\bf x} \|) \, \ud \tilde {\bf x} + O(\epsilon^{d+\delta}).
\end{align}
Inserting \cref{Bcfinal} into \cref{sfp_int}, we find that the marginal density function satisfies the following nonlinear Fokker--Planck equation
\begin{subequations}
\label{sfpNfinal_close}
\begin{align}
\label{sfpNfinal_close_eq}
\frac{\partial p}{\partial t} = \nabla_{
 {\bf  x}_1} \cdot   \left \{ \big [1 + \overline \alpha_u (N-1) \epsilon^d p \big] {\nabla}_{{\bf  x}_1}  p -  {\bf  f}({\bf  x}_1) p \right \},
\end{align}
where the nonlinear coefficient is given by
\begin{equation}
\label{alphaV_close}
\overline \alpha_{u} =  \int_{\mathbb R^d}   u( \epsilon \| {\bf x}  \| ) \, \ud  {\bf x}.
\end{equation}
\end{subequations}
We will refer to \eqref{sfpNfinal_close}, in which the mean-field closure has been combined with the assumption of a short-range potential, as the localized MFA or LMFA. 
As we shall see later, the MFA or LMFA are not defined for many commonly used short-range repulsive potentials. If the integral in \cref{alphaV_close} does not exist because of the behavior at infinity then this is an indication that the potentials is too long range for the localization performed above to be valid and the full MFA integral needs to be retained. However, if the integral in \cref{alphaV_close} diverges because of the behavior at the origin then the MFA itself will diverge, that is, the integral in \cref{Bcclosed} will not exist. Examples of the inappropriate use of MFA for such short range potentials exist in the literature \cite{Horng:2012io}. 

\subsection{Closure at the pair correlation function} \label{sec:felderhof}
A more elaborated closure is suggested by Felderhof \cite{Felderhof:1978vn}. His derivation considers a more general context of interacting Brownian particles suspended in a fluid, including hydrodynamic interactions. His analysis is valid for zero external force, ${\bf f} \equiv 0$, and is based on the thermodynamic limit (in which the number of particles $N$ and the system volume $V$ tend to infinity, with the number density $N/V = \rho_0$ fixed). Because of this, instead of working with probability densities, it is convenient to switch to number densities: $\rho ({\bf x}_1,t) : = \rho_0 p({\bf x}_1,t)$ and $Q ({\bf x}_1,{\bf x}_2, t) := \rho_0^2 P_2({\bf x}_1,{\bf x}_2, t)$. In what follows we outline his derivation for hard spheres ignoring hydrodynamic interactions. The equation for the one-particle number density $\rho({\bf x}_1,t)$ is
\begin{equation}
\label{felde1}
\frac{\partial \rho}{\partial t}  = \nabla_{{\bf  x}_1}  \cdot \left( \nabla_{{\bf  x}_1} \,  \rho  + \int Q({\bf x}_1, {\bf x}_2, t) \nabla_{{\bf  x}_1} u(r) \, \ud {\bf x}_2  \right),
\end{equation}
where $r = \|{\bf x}_1- {\bf x}_2 \|$ is the interparticle distance and $Q({\bf x}_1, {\bf x}_2, t)$ (the two-particle number density) satisfies, to lowest order in $\rho_0$, 
\begin{equation}
\label{felde2}
\frac{\partial Q}{\partial t} = \nabla_{{\bf  x}_1}  \cdot \left[ \nabla_{{\bf  x}_1} \,  Q + Q \nabla_{{\bf  x}_1} u(r) \right] + \nabla_{{\bf  x}_2}  \cdot \left[ \nabla_{{\bf  x}_2} \,  Q + Q \nabla_{{\bf  x}_2} u(r) \right].
\end{equation}
Equations \cref{felde1} and \cref{felde2} have the following time-independent equilibrium solutions
\begin{equation}
\label{felde_ss}
\rho_s({\bf x}_1) = \rho_0, \qquad Q_{s} ({\bf x}_1,{\bf x}_2) = \rho_0^2 \,  g_0 ( \|{\bf x}_1- {\bf x}_2 \|),
\end{equation}
where $\rho_0$ is constant  and $g_0$ is the pair correlation function
\begin{equation}
\label{felde_corr}
g_0(r) = e^{-u(r)}.
\end{equation}
Felderhof then looks for a linearized solution around the equilibrium values \cref{felde_ss}, by making the ansatz
\begin{equation}
\label{felde_ansatz}
Q({\bf x}_1,{\bf x}_2, t) = \rho({\bf x}_1,t) \rho ({\bf x}_2,t) g({\bf x}_1,{\bf x}_2, t),
\end{equation}
and considering the deviations $\rho_1$ and $g_1$,
\begin{equation}
\label{felde_linia}
\rho({\bf x}_1,t)  = \rho_0 + \rho_1({\bf x}_1,t), \qquad g({\bf x}_1,{\bf x}_2, t) = g_0(r\\)  + g_1({\bf x}_1,{\bf x}_2, t).
\end{equation}
Then, to terms linear in $\rho_1$ and $g_1$,  \cref{felde_ansatz} becomes
\begin{equation}
\label{felder_linia2}
Q({\bf x}_1,{\bf x}_2, t)  \approx \rho_0^2\,  g_0(r) +  \rho_1({\bf x}_1,t) \rho_0\,  g_0(r) + \rho_0 \rho_1 ({\bf x}_2,t) g_0(r) + \rho_0^2\,  g_1({\bf x}_1,{\bf x}_2, t).
\end{equation}
Substituting in \cref{felde1} and linearizing gives 
\begin{multline}
\label{felde3}
\frac{\partial \rho_1}{\partial t} = \nabla_{{\bf  x}_1}  \cdot \Big[   \nabla_{{\bf  x}_1}  \rho_1  + \rho_0 \int g_0(r) \rho_1({\bf x}_2, t)   \nabla_{{\bf  x}_1} u(r) \, \ud {\bf x}_2 \\
+ \rho_0^2 \int g_1({\bf x}_1,{\bf x}_2, t) \nabla_{{\bf  x}_1} u(r) \, \ud {\bf x}_2 \Big].
\end{multline}
At this stage, Felderhof supposes that perturbations from equilibrium \cref{felde_ss} are small so that $\nabla_{{\bf  x}_1}  \rho_1({\bf x}_1,t) \approx \nabla_{{\bf  x}_2}  \rho_1({\bf x}_2,t)$ and the pair correlation function is at its equilibrium value, that is, $g({\bf x}_1,{\bf x}_2, t) \approx g_0(r)$. Then \cref{felde3} simplifies to
\begin{align}
\label{felde4}
\frac{\partial \rho_1}{\partial t}  = \nabla_{{\bf  x}_1}  \cdot  \left [ \nabla_{{\bf  x}_1}  \rho_1  + \rho_0 \! \int \! g_0(r) \rho_1({\bf x}_2, t)   \nabla_{{\bf  x}_1} u(r) \, \ud {\bf x}_2 \, \right].
\end{align}
Now, using \cref{felde_corr} and expanding $\rho_1({\bf x}_2,t)$ about ${\bf x}_1$ and keeping only the first non-vanishing term gives \cite{Felderhof:1978vn}
\begin{align}
\label{felde5}
\frac{\partial \rho_1}{\partial t}  = \nabla_{{\bf  x}_1}  \cdot \left[ (1+  \alpha_u \epsilon^d \rho_0) \nabla_{{\bf  x}_1}  \rho_1   \right],
\end{align}
where $\alpha_u = \int_{\mathbb R^d}  \left ( 1 - e^{- u(\epsilon \| {\bf x} \|)} \right)  \ud  {\bf x}.$ Note that this is the evolution equation for the perturbation $\rho_1$ from the uniform equilibrium $\rho_0$ (valid with ${\bf f}= 0$).

\subsection{Kirkwood closure}

As we will see later in the results section, the MFA can only provide an adequate approximation for relatively soft interaction potentials and low densities.  An alternative closure approximation is based on the Kirkwood superposition approximation (KSA) \cite{Kirkwood:1935is}, and consists of truncating the hierarchy at the two-particle density function. To this end, we consider the equation satisfied by $P_2({\bf  x}_1, {\bf  x}_2, t)$ by integrating the $N$-particle Fokker--Planck equation \cref{sfp_eq} over ${\bf x}_3, \dots, {\bf x}_N$, applying the divergence theorem and relabelling particles as before to obtain
\begin{multline} \label{P2eqhier}
\frac{\partial P_2}{\partial t}
=   \nabla_{{\bf  x}_1} \cdot \left[ \nabla_{{\bf x}_1}   P_2 -  {\bf  f}({\bf  x}_1)  P_2 + \nabla_{ \bfx_1} u( \| \bfx_1 - \bfx_2 \| ) P_2 +  {\bf G}_2(\bfx_1, \bfx_2, t) 
  \right]  \\
 +   \nabla_{{\bf  x}_2} \cdot \left[ \nabla_{{\bf x}_2}   P_2 -  {\bf  f}({\bf  x}_2)  P_2 + \nabla_{ \bfx_2} u( \| \bfx_1 - \bfx_2 \| ) P_2 +  {\bf G}_2(\bfx_2, \bfx_1, t)  \right], 
\end{multline}
where
\begin{equation} \label{G2}
{\bf G}_2(\bfx_1, \bfx_2, t)  =  (N-2) \int_\Omega \nabla_{ \bfx_1} u( \| \bfx_1 - \bfx_3 \| ) P_3( \bfx_1,\bfx_2,\bfx_3,t)  \, \ud \bfx_3,
\end{equation}
and 
\begin{equation} 
P_3( {\bf x}_1, {\bf x}_2, \bfx_3, t) = \int_{\Omega^{N-3}} P(\vec x, t) \,  \ud {\bf x}_4 \cdots \ud {\bf x}_N
\end{equation}
is the three-particle density function. We note that in writing \cref{G2} we are using that $P_3$ is invariant to particle relabelling. 
The KSA then approximates the three-particle density function as
\begin{equation}
\label{kirkwood}
P_3({\bf x}_1,{\bf x}_2,{\bf x}_3,t) = \frac{ P_2 ({\bf x}_1,{\bf x}_2,t) P_2 ({\bf x}_1,{\bf x}_3,t) P_2 ({\bf x}_2,{\bf x}_3,t) } {p ({\bf x}_1,t) p ({\bf x}_2,t) p ({\bf x}_3,t) },
\end{equation}
where $p$ and $P_2$ are the one- and two-particle density functions, respectively. Inserting \cref{kirkwood} into \cref{G2} one can then solve the coupled system \cref{sfp_int} and \cref{P2eqhier} for $p$ and $P_2$. 

The KSA closure has been the basis of many subsequent closure approximations, and it can be derived as the maximum entropy closure in the thermodynamic limit \cite{Singer:2004ec}. Because it is thought to be superior to the MFA at high densities, it has been used in several biological applications such as on-lattice birth-death-movement processes with size-exclusion \cite{Baker:2010is} and off-lattice cell motility processes with soft interactions \cite{Markham:2013tv,Middleton:2014fa}.  Middleton, Fleck and Grima \cite{Middleton:2014fa} consider a system of Brownian particles evolving according to \cref{ssde} in one dimension interacting via a Morse potential (see \cref{pair_MO}) and compare the KSA closure to the MFA closure and simulations of the stochastic system. Markham et al. \cite{Markham:2013tv} use the KSA on a more general individual-based model, where the random jumps of particles are not Gaussian but depend on the positions of all particles, resulting in multiplicative noise. Berlyand, Jabin and Potomkin \cite{Berlyand:2016fu} use a variant of the KSA closure, approximating either $P_2( {\bf x}_1, {\bf x}_3, t)$ or $P_2( {\bf x}_2, {\bf x}_3, t) $  in \cref{kirkwood} by their corresponding mean-field approximation $P_2( {\bf x}_i, {\bf x}_j, t) = p( {\bf x}_i, t) p( {\bf x}_j, t)$, for a system of interacting deterministic particles. 

It is worth noting that the KSA model is computationally expensive and complicated to solve, especially if the interaction potential $u$ is short ranged, requiring a fine discretization. For example, in 3 dimensions one must solve a 6-dimensional problem which, once discretized, involves a full discretization matrix because of the convolution terms ${\bf G}$ and ${\bf G}_2$. As we shall see in \cref{sec:numerics}, even in one physical dimensional the KSA model is rather complicated to solve.

\section{Matched asymptotic expansions} \label{sec:soft_mae}

In this section we consider an approach based on matched asymptotic expansions (MAE) to obtain a closed equation for the one-particle density $p$ that is valid for short-range interaction potentials, and that is computationally practical to solve even in two or three dimensions. 

We go back to the evolution equation \cref{sfp_int} for the one-particle density $p$. Assuming that the pair potential $u$ is localized near ${\bf x}_1$, we can determine $\bf G$ in \cref{Balltrue} using MAE. To do so, we first must obtain an expression for $P_2$. 

For low-concentration solutions with short-range interactions, three-particle (and higher) interactions are negligible compared to two-particle interactions: when two particles are close to each other, the probability of a third particle being nearby is so small that it can be ignored. Mathematically, this means that the two-particle probability density $P_2( {\bf x}_1, {\bf x}_2, t)$ is governed by the dynamics of particles 1 and 2 only, independently of the remaining $N-2$ particles. 
In other words, the terms ${\bf G}_2$ in \cref{P2eqhier} are negligible and the equation for $P_2$ reduces to 
\begin{multline}
\label{sfp2_eq} 
\frac{\partial P_2}{\partial t}
=   \nabla_{{\bf  x}_1} \cdot \left[ \nabla_{{\bf x}_1}   P_2 -  {\bf  f}({\bf  x}_1)  P_2 + \nabla_{ \bfx_1} u( \| \bfx_1 - \bfx_2 \| ) P_2 
  \right]  \\
 +   \nabla_{{\bf  x}_2} \cdot \left[ \nabla_{{\bf x}_2}   P_2 -  {\bf  f}({\bf  x}_2)  P_2 + \nabla_{ \bfx_2} u( \| \bfx_1 - \bfx_2 \| ) P_2   \right], 
\end{multline}
for  $({\bf  x}_1, {\bf  x}_2) \in \Omega ^2$, complemented with periodic boundary conditions on $\partial \Omega^2$. Note that in approximating $P_2$ in this way it is no longer true that $p(\bfx_1,t) = \int P_2(\bfx_1, \bfx_2,t) \ud \bfx_2$, so that $p$ and $P_2$ need to be solved for as a coupled system. This equation is basically \cref{felde2} with an added external force. Essentially our MAE approach aims to solve \cref{felde1,felde2} systematically asymptotically rather than through Felderhof's linearization and approximations (see \cref{sec:felderhof}).

\subsection{Inner and outer regions} \label{sec:soft-inner-outer}

By assumption, the pair interaction potential $u(r)$ is negligible everywhere except when the interparticle distance $r$ is of order $\epsilon$. Therefore, we suppose that when two particles are far apart ($\|{\bf  x}_1 -
{\bf  x}_2\|\gg \epsilon$) they are independent, whereas when
they are close to each other ($\|{\bf  x}_1 - {\bf  x}_2\|
\sim \epsilon$) they are correlated. We designate these two regions of configuration space the outer region and inner region, respectively. 
 
In the outer region  we define $P_\text{o}({\bf  x}_1, {\bf  x}_2, t) = P_2( {\bf  x}_1, {\bf  x}_2, t)$. By independence, we have that\footnote{Independence only tells us that $P_\text{o}({\bf  x}_1, {\bf  x}_2, t) \sim q({\bf  x}_1, t) q({\bf  x}_2, t)$ for some function $q$ but the normalization condition on $P$ implies $p = q + O(\epsilon)$.}
\begin{equation} 
\label{sindep}
P_\text{o}({\bf  x}_1, {\bf  x}_2, t) = p({\bf  x}_1, t) p({\bf  x}_2, t) + \epsilon P_\text{o}^{(1)}({\bf  x}_1, {\bf  x}_2, t) + \cdots, 
\end{equation}
for some function $P_\text{o}^{(1)}$. 

In the inner region, we set  ${\bf  x}_1 = \tilde {\bf x}_1$ and ${\bf  x}_2 =
\tilde {\bf x}_1 + \epsilon \tilde {\bf x}$ and define $\tilde P (\tilde {\bf x}_1, \tilde {\bf x}, t) = P_2( {\bf x}_1,  {\bf x}_2 , t)$ and $\tilde u (\| \tilde {\bf x} \| ) = u(\|{\bf x}_1-  {\bf x}_2\|)$. Rewriting \cref{sfp2_eq} in terms of the inner coordinates gives 
\begin{multline}
\label{sfp3_eq} 
\epsilon^2 \frac{\partial \tilde P}{\partial t}
=   2 \nabla_{\tilde {\bf  x}}  \cdot  \left[ \nabla_{\tilde {\bf  x}}  \tilde P + \nabla_{\tilde {\bf  x}} \tilde u( \| \tilde {\bf  x} \| ) \tilde P \right] + \epsilon  {\nabla} _{\tilde {\bf x}} \cdot \left\{ \left[ {\bf  f}(\tilde {\bf x}_1) -  {\bf   f}(\tilde {\bf x}_1 + \epsilon \tilde {\bf x}) \right] \tilde P \right \}\\ -  \epsilon  \nabla_{\tilde { {\bf  x}}_1}  \cdot  \left[ 2 \nabla_{\tilde {\bf  x}}  \tilde P + \nabla_{\tilde {\bf  x}}  \tilde u (  \| \tilde {\bf  x} \| ) \tilde P  \right] + \epsilon^2 \nabla_{\tilde {\bf x}_1}^2 \, \tilde P - \epsilon ^2  {\bo\nabla} _{\tilde {\bf x}_1} \cdot \left[ {\bf  f}(\tilde {\bf x}_1)  \tilde P \right ]. 
\end{multline}
The inner solution $\tilde P$ must match with the outer solution $P_\text{o}$ as $\|\tilde {\bf x} \| \rightarrow \infty$.  Expanding $P_\text{o}$ in terms of the inner variables gives (omitting the time variable for ease of notation)
\begin{align}
\label{4:bc_match}
\begin{aligned}
P_\text{o} ( {\bf  x}_1, {\bf x}_2) &\sim p(\tilde {\bf x}_1) p(\tilde {\bf x}_1 + \epsilon \tilde {\bf  x}) + \epsilon P_\text{o}^{(1)}(\tilde {\bf  x}_1, \tilde {\bf x}_1 + \epsilon \tilde {\bf  x}) \\
& \sim  p^2 (\tilde {\bf x}_1) +  \epsilon p(\tilde {\bf  x}_1) \, \tilde {\bf x} \cdot \nabla _{ \tilde  {\bf x}_1} p( \tilde {\bf x}_1) + \epsilon P_\text{o}^{(1)}(\tilde {\bf  x}_1, \tilde {\bf x}_1)  \qquad \textrm{as} \qquad \|\tilde {\bf x} \|\rightarrow \infty.
\end{aligned}
\end{align}

We look for a solution of \cref{sfp3_eq} matching with \cref{4:bc_match}  as $\|\tilde {\bf x} \|\rightarrow \infty$ of the form
$\tilde P  \sim \tilde P^{(0)} + \epsilon\tilde P^{(1)} +  \cdots$.
The leading-order inner problem is 
\begin{align} \label{so1}
0 = 2 \nabla_{\tilde {\bf  x}} \cdot  \left[ \nabla_{\tilde {\bf  x}}  \tilde P^{(0)} + \nabla_{\tilde {\bf  x}} \tilde u( \| \tilde {\bf  x} \| ) \tilde P^{(0)} \right],  \qquad
\tilde P^{(0)} \sim p^2 (\tilde {\bo {\mathrm x}}_1) \quad \textrm{as} \quad \|\tilde {\bo {\mathrm x}}\|\to \infty,
\end{align}
with solution 
\begin{equation}
\label{stwo_order0_sol}
\tilde P^{(0)} =  p^2 (\tilde {\bo {\mathrm x}}_1) \, e^{-\tilde u( \| \tilde {\bf x} \| )}.
\end{equation} 

The $O(\epsilon)$ problem reads
\begin{subequations}
\label{stwo_order1}
\begin{align}
\label{stwo_order1_eq}
0&=2 \nabla_{\tilde {\bf  x}}  \cdot  \left[\nabla_{\tilde {\bf  x}}  \tilde P^{(1)} + \nabla_{\tilde {\bf  x}} \tilde u( \| \tilde {\bf  x} \|) \tilde P^{(1)} \right]  -    \nabla_{\tilde { {\bf  x}}_1}  \cdot  \left[ 2 \nabla_{\tilde {\bf  x}}  \tilde P^{(0)} + \nabla_{\tilde {\bf  x}}  \tilde u( \| \tilde {\bf  x} \|) \tilde P^{(0)}  \right], 
\\
\label{stwo_order1_bc1}
\tilde P^{(1)} &\sim  p(\tilde {
  {\bf  x}}_1) \, \tilde {\bf x} \cdot \nabla _{ \tilde
  {\bf x}_1} p( \tilde {\bf x}_1) + P_\text{o}^{(1)}(\tilde {\bf  x}_1, \tilde {\bf x}_1)  \qquad \textrm{as} \qquad \|\tilde {\bo {\mathrm x}}\|\sim \infty.
\end{align}
\end{subequations}
Using \cref{stwo_order0_sol}, we can rearrange \cref{stwo_order1_eq} to give
\begin{equation} 
\label{stwo_order1_eqb}
\nabla_{\tilde {\bf  x}}  \cdot  \left[ \nabla_{\tilde {\bf  x}}  \tilde P^{(1)} + \nabla_{\tilde {\bf  x}} \tilde u( \| \tilde {\bf  x} \|) \tilde P^{(1)}  - \tfrac{1}{2} \nabla_{\tilde { {\bf  x}}_1}  \tilde P^{(0)} \right]  = 0.
\end{equation}
Solving \cref{stwo_order1_eqb} together with \cref{stwo_order1_bc1} gives
\begin{equation}
\label{stwo_order1_sol}
\tilde P^{(1)} =  \left[ p (\tilde {\bo {\mathrm x}}_1)   \tilde {\bf x} \cdot \nabla _{ \tilde
  {\bf x}_1} p( \tilde {\bf x}_1) + P_\text{o}^{(1)}(\tilde {\bf  x}_1, \tilde {\bf x}_1) \right] e^{-\tilde u( \| \tilde {\bf x} \|)}.
\end{equation} 
Thus we find that the inner region solution is, to $O(\epsilon)$,
\begin{align}
\label{s_innersol}
\tilde P \sim \left[  p^2 (\tilde {\bo {\mathrm x}}_1,t) + \epsilon  p (\tilde {\bo {\mathrm x}}_1,t) \, \tilde {\bf x} \cdot \nabla _{ \tilde
  {\bf x}_1} p( \tilde {\bf x}_1,t) + \epsilon P_\text{o}^{(1)}(\tilde {\bf  x}_1, \tilde {\bf x}_1,t) + \cdots \right] e^{-\tilde u( \| \tilde {\bf x} \|)}.
\end{align}

\subsection{Interaction integral} \label{sec:soft-interactionintegral}
Now we go back to the interaction integral ${\bf G}({\bf x}_1)$ in \cref{Balltrue}. Because of the short-range nature of the potential $u$, the main contribution to this integral is from the inner region. Therefore, we will use the inner solution \cref{s_innersol} to evaluate it. 

First we split the integration volume $\Omega$ for ${\bf x}_2$ into the inner and the outer regions defined in the previous section. Although there is no sharp boundary between the inner and outer regions, it is convenient to introduce an intermediate radius $\delta$, with $\epsilon \ll \delta \ll 1$, which divides the two regions. Then the inner region is $\Omega_\text{i}({\bf x}_1) = \{ {\bf x}_2 \in \Omega \ : \ \| {\bf x}_2- {\bf x}_1 \| <\delta \}$ and the outer region is the complimentary set $\Omega_\text{o}({\bf x}_1) = \Omega \setminus \Omega_\text{i}({\bf x}_1)$. The dominant contribution to \cref{Balltrue} is then
\begin{align}
\label{svecB}
\begin{aligned}
{\bf G} ({\bf x}_1, t) &= (N-1) \int_{ \Omega_\text{i}({\bf x}_1) }  P({\bf x}_1, {\bf x}_2, t) \nabla_{{\bf x}_1} \! u( \| {\bf x}_1- {\bf x}_2 \|)  \, \ud {\bf x}_2 \\
 &= -(N-1) \epsilon^{d-1}\! \! \int_{\| \tilde {\bf x} \| < \delta/\epsilon }\! \tilde P( {\bf x}_1, \tilde {\bf x}, t)  \nabla_{\tilde {\bf x}} \tilde u( \| \tilde {\bf x} \|)  \, \ud \tilde {\bf x} \\
 &\sim  (N-1) \epsilon^{d-1}  p ( {\bf x}_1,t) \! \! \int_{\| \tilde {\bf x} \| < \delta/\epsilon } \!  \bigg\{ p ( {\bf x}_1,t) + \epsilon \tilde {\bf x} \cdot \nabla _{   {\bf x}_1} p(  {\bf x}_1,t) \\ & \hspace{5.7cm} +  \epsilon P_\text{o}^{(1)}( {\bf  x}_1,  {\bf x}_1)   \bigg \} 
\nabla_{\tilde {\bf x}}   e^{-\tilde u( \| \tilde {\bf x} \|)} \, \ud \tilde {\bf x}.
\end{aligned}
\end{align}
The first and third terms of the integral vanish using the divergence theorem and that $\tilde u$ is a radial function of $\|\tilde {\bf x}\|$. Integration by parts on the second component gives
\begin{multline*}
\int_{\| \tilde {\bf x} \| < \delta/\epsilon } \! \left[ \tilde {\bf x} \cdot \nabla _{  {\bf x}_1} p( {\bf x}_1,t) \right]  
\nabla_{\tilde {\bf x}}   e^{-\tilde u( \| \tilde {\bf x} \|)} \, \ud \tilde {\bf x}\\  \sim \nabla _{  {\bf x}_1} p( {\bf x}_1,t)\!  \bigg[ \tfrac{2(d-1)\pi}{d} \left ( \tfrac{\delta}{\epsilon} \right)^d - \int_{\| \tilde {\bf x} \| <  \delta/\epsilon }   e^{-\tilde u( \| \tilde {\bf x} \|)} \, \ud \tilde {\bf x}\bigg ],
\end{multline*}
where we have used that $e^{-\tilde u( \| \tilde {\bf x} \|)} \approx 1$ at $\|\tilde {\bf x}\| = \delta/\epsilon$. Finally, rewriting the first term above as a volume integral,\footnote{The volume of a $d$-dimensional sphere of radius $\delta/\epsilon$  is equal to \mbox{$2\pi (d-1)/d (\delta/\epsilon)^d$}.}   \cref{svecB} becomes
\begin{align*}
{\bf G}({\bf x}_1, t)  \sim  (N-1) \epsilon^d p( {\bf x}_1,t) \nabla _{ {\bf x}_1} p({\bf x}_1,t) \int_{\| \tilde {\bf x} \| <  \delta/\epsilon } \left ( 1 - e^{-\tilde u( \| \tilde {\bf x} \|)} \right)  \ud \tilde {\bf x}.
\end{align*}
Since $1 - e^{-\tilde u( \| \tilde {\bf x} \|)}$ decays at infinity, we can extend the domain of integration to the entire $\mathbb R^d$ introducing only lower order errors. Therefore we can write
\begin{align}
\label{BMAE}
{\bf G}( {\bf x}_1, t)  \sim  \alpha_u (N-1) \epsilon^d p( {\bf x}_1,t) \nabla _{ {\bf x}_1} p(  {\bf x}_1,t),
\end{align}
with
\begin{equation*} 
\alpha_u =  \int_{\mathbb R^d}  \left ( 1 - e^{-\tilde u( \| \tilde {\bf x} \|)} \right)  \ud \tilde {\bf x} = \int_{\mathbb R^d}  \left ( 1 - e^{- u( \epsilon \|  {\bf x} \|)} \right)  \ud {\bf x}.
\end{equation*}
Note that we obtain the same coefficient $\alpha_u$ that appeared in equation \cref{felde5} using the closure at the pair correlation function. 

\subsection{Reduced Fokker--Planck equation for soft spheres} \label{sec:soft_FP}

Combining \cref{BMAE} with  \cref{sfp_int} we find that, to $O(\epsilon^d)$,
\begin{subequations}
\label{sfpNfinal}
\begin{align}
\label{sfpNfinal_eq}
\frac{\partial p}{\partial t}=  \nabla_{
 {\bf  x}_1} \cdot   \left \{ \big [1 +  \alpha_u (N-1) \epsilon^d p \big] {\nabla}_{{\bf  x}_1}  p -  {\bf  f}({\bf  x}_1) p \right \},
\end{align}
where 
\begin{equation} 
\label{alphaV}
\alpha_u = \int_{\mathbb R^d}  \left ( 1 - e^{- u(\epsilon \| {\bf x} \|)} \right)  \ud  {\bf x}.
\end{equation}
\end{subequations}
Therefore, we find that the MAE method yields the same type of equation as the LMFA (see \cref{sfpNfinal_close}), but with a different coefficient in the nonlinear diffusion term. Expanding the exponential in \cref{alphaV} for small $u$ gives 
\begin{align*}
\alpha_u = \int_{\mathbb R^d}  \left ( 1 - e^{- u( \epsilon \| {\bf x} \|)} \right)  \ud  {\bf x} = \int_{\mathbb R^d}  u(\epsilon \| {\bf x}\|)\, \ud  {\bf x} - \frac{1}{2}  \int_{\mathbb R^d} u^2(\epsilon \|{\bf x}\|) \, \ud  {\bf x} + \cdots, 
\end{align*}
that is, the LMFA closure is the leading contribution of the potential, provided that $u(\epsilon \| {\bf x}\|)$ is small. However, as we will see in \cref{sec:results}, this is not always true. We also see that the equation obtained by Felderhof \cite{Felderhof:1978vn}, \cref{felde5} is the linearized version of \cref{sfpNfinal_eq} after taking  $N \to \infty$ and setting the external force to zero. 

The coefficient $\alpha_u$ can be related to basic concepts from statistical mechanics. Namely, the integrand in \cref{alphaV}  is the negative of the \emph{total correlation function} $h(r) = g(r)-1$, where $g(r) = \exp[-u(r)]$ is the low-density limit of the pair correlation function, the so-called Boltzmann factor of the pair potential \cite[\textsection2.6]{Hansen:2006uv}. 

\subsection{Two-particle density function via MAE}

In contract to the MFA, using MAEs we have obtained an approximation of the two-particle density function $P_2(\bfx_1, \bfx_2, t)$ in two regions of the configuration space, the outer and the inner regions, defined according to the separation between the two particles. It is convenient to have a uniformly valid expansion in the whole space--so that for example we can plot and compare it against  simulations of the stochastic particle system. This can be done by constructing a so-called composite expansion, consisting of the inner expansion \cref{s_innersol} plus the outer expansion \cref{sindep} minus the common part \cite[Chapter 5]{Hinch:1991go}. The result is
\begin{equation} \label{MAE_p2}
P_2({\bf x}_1, {\bf x}_2, t) \sim p({\bf x}_1,t) p({\bf x}_2,t) e^{-u( \| {\bf x}_1 - {\bf x}_2 \| )}.
\end{equation}

\section{Results} 
\label{sec:results}

\subsection{The hard-sphere potential} \label{sec:softhard_pot}

So far we have assumed that the set of Brownian particles interact via a soft pair potential $u(r)$. However, the resulting reduced Fokker--Planck equation \cref{sfpNfinal} obtained via MAE can be used to model a system of hard-core interacting particles of diameter $\epsilon$. 
In particular, the model \cref{sfpNfinal} for soft spheres has exactly the same structure as the counterpart model for hard spheres derived in \cite{Bruna:2012cg}. Inserting the hard-sphere pair potential
\begin{equation}
\label{pairpotentialHS}
u_\text{HS}(r ) = \left \{ 
\begin{aligned}  
&\infty \quad &r \le \epsilon, \\
&0 \quad &r >\epsilon,
 \end{aligned} \right.
\end{equation}
into the nonlinear diffusion coefficient in \cref{alphaV} gives
\begin{equation}
\label{salphaHS}
\alpha_\text{HS} =  \int_{\| {\bf x}\|<1} \ud  {\bf x},
\end{equation}
that is, $\alpha_\text{HS} = 2$ for $d=1$, $\alpha_\text{HS} = \pi$ for $d=2$, and $\alpha = 4\pi /3$ for $d=3$. This is in agreement with our previous work where the reduced model was specifically derived for a system of hard spheres \cite{Bruna:2012cg,Bruna:2013ku}. For hard spheres the configuration space has holes due to the illegal configurations (associated with infinite energy) and the derivation is slightly different.
We note that the MFA for hard spheres does not work and that, in particular, the coefficient $\overline \alpha_\text{HS}$ in \cref{alphaV_close} is not defined.

Because the MAE models for soft and hard spheres coincide, for every potential $u$ we can find an effective hard-sphere diameter $\epsilon_\text{eff}$ such that the continuum models of both systems -- the soft-sphere system with pair potential $u(r)$ and the hard-sphere system with diameter $\epsilon_\text{eff}$ -- are equivalent. In other words, a characterization of a given soft potential $u$ is to find $\epsilon_\text{eff}$ such that 
\begin{equation}
\label{epeff0}
\alpha_u \epsilon^d = \alpha_\text{HS} \, \epsilon_\text{eff}^d. 
\end{equation}
where $\alpha_u$ is given in \cref{alphaV} and $\alpha_\text{HS} $ given in \cref{salphaHS}. Rearranging, we find that
\begin{equation}
\label{epeff}
\left(\frac{\epsilon_\text{eff}}{\epsilon}\right)^d = d  \int_0^\infty  \left ( 1 - e^{- u(\epsilon r)} \right) r^{d-1} \, \ud r.
\end{equation}

This idea of finding the effective hard-sphere diameter associated to a soft-sphere system, known as the effective hard sphere diameter method, has been widely used to calculate both equilibrium and transport properties \cite[\textsection9.3.3]{Mulero:2012vc}. The reason this method is appealing is that it allows us to ``translate'' a general system of interacting soft spheres (whose properties may have not been studied before) to the widely studied hard-sphere system, on which most theories are based. 
Moreover, the derived model \cref{sfpNfinal} implies that, as far as the population-level dynamics are concerned, soft interactions may be incorporated into the effective hard particle model by adjusting the hard-sphere diameter with \cref{epeff}. This is provided $\alpha_u$ in \cref{alphaV} is well-defined and positive. In contrast, if $\alpha_u$ does not exist (or is negative), the MAE approach breaks down (or may become unstable) for the given pair potential $u$. Even this is instructive: it means that the potential is not decaying at infinity fast enough to be incorporated into a population-level equation of the form \cref{sfpNfinal_eq} and that the MFA is preferrable.  

Another application of the effective hard-sphere diameter is to provide an effective volume fraction for the system of soft spheres: using $\epsilon_\text{eff}$, one can define the effective volume fraction of soft spheres and use it to check whether the ``low-volume fraction'' condition holds. 
 
\subsection{Comparison between MAE and LMFA} 
\label{sec:comparison}

\begin{subequations}
\label{soft_potentials}
The simplest repulsive pair potential is the soft-sphere (SS) potential, which assumes the form
\begin{equation} 
\label{pairpotentialSS}
u_{\text{SS}} (r) =  (\epsilon/r)^\nu,
 \end{equation}
where $\epsilon$ is a measure of the range of the interaction and $\nu$ is  the hardness parameter which characterizes the particles (the softness parameter is defined
as its inverse, $1/\nu$). We note that for $\nu =1$, the SS potential corresponds to the Coulomb interaction ($\nu =1$)  \cite[Chapter 9]{Mulero:2012vc}. Other common purely repulsive potentials include the exponential (EX) potential \cite[Chapter 7]{Israelachvili:2011ug}
\begin{equation}
\label{pairpotentialEX}
u_{\text{EX}} (r) = e^{-r/\epsilon},
\end{equation}
and the repulsive Yukawa (YU) potential 
\begin{equation}
\label{pairpotentialYU}
u_\text{YU} (r) = \frac{\epsilon}{r} e^{-r/\epsilon}.
\end{equation}
This potential, also known as screened Coulomb, is used to describe elementary particles, small charged ``dust'' grains observed in plasma environments, and suspensions of charge-stabilized colloids \cite{Hynninen:2003hr}. 

To model some physical systems it is convenient to incorporate an attractive part to the repulsive pair potential. The most common situation is that particles repel each other in the short range and attract each other in a longer range. For example, the SS potential in \cref{pairpotentialSS} may be generalized to power-law repulsive--attractive potentials of the form $u(r) = (\epsilon/r)^a - C(\epsilon/r)^b$, with $a>b$ and $C>0$, known as the Mie potentials. The most famous example of this class of potentials is the Lennard--Jones potential,
\begin{equation}
\label{pair_LJ}
u_\text{LJ} (r) =  \left( \frac{\epsilon}{r} \right)^{12} -  \left( \frac{\epsilon}{r} \right)^{6}.
\end{equation}
Another common repulsive--attractive potential is the Morse (MO) potential
\begin{equation}
\label{pair_MO}
u_\text{MO}(r) = e^{-r/\epsilon} - \frac{1}{C} e^{-lr/\epsilon},
\end{equation}
where $C$ and $l$ are, respectively, the relative strength and relative lengthscale of the repulsion to attraction. The most relevant situations for biological applications are given for $C>1$ and $l<1$, which correspond to short-range repulsion and weak long-range attraction \cite{Carrillo:2010tx}. 
\end{subequations}
\cref{fig:potentials_soft} shows examples of all the interaction potentials above. 
\def \scc {0.75}
\def \scl {1.0}
\begin{figure}[ht]
\unitlength=1cm
\begin{center}
\psfragscanon
\psfrag{r}[][][\scl]{$r/\epsilon$}  \psfrag{u}[][][\scl][0]{$u(r)$} 
\psfrag{datadatadata1}[][][\scc]{$u_\text{HS}$, \cref{pairpotentialHS}} \psfrag{datadatadata2}[][][\scc]{$u_\text{SS}$, \cref{pairpotentialSS}}
\psfrag{datadatadata3}[][][\scc]{$u_\text{EX}$, \cref{pairpotentialEX}} \psfrag{datadatadata4}[][][\scc]{$u_\text{YU}$, \cref{pairpotentialYU}}      
\psfrag{datadatadata5}[][][\scc]{$u_\text{LJ}$,  \cref{pair_LJ}} \psfrag{datadatadata6}[][][\scc]{$u_\text{MO}$, \cref{pair_MO}}    
\includegraphics[width = .6\linewidth]{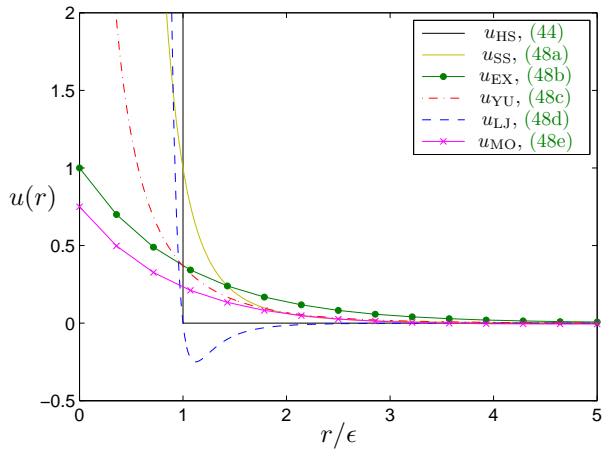}
\caption{Hard-sphere potential \cref{pairpotentialHS} and soft potentials \cref{soft_potentials}. Potential SS corresponding to $\nu = 4$ and Morse potential for $C = 4$ and $l=0.6$. }
\label{fig:potentials_soft}
\end{center}
\end{figure}

A simple way to compare between approaches for short-range interaction potentials is to consider the nonlinear diffusion term using either MAE or LMFA. Their respective coefficients $\alpha_u$ in \cref{alphaV}  and $\overline \alpha_u$ in \cref{alphaV_close} for the potentials above are shown in \cref{table:comparison}. Since the behaviour of the  integrals in $\alpha_u$ and $\overline \alpha_u$ at infinity is the same, the LMFA fails for any case in which the MAE fails. LMFA may fail also because of a singularity at the origin for which MAE is valid, as seen in some examples of \cref{table:comparison}. This is because the strongly repulsive short-range part of the potential results in correlations which violate the  MFA that particles may be considered independent. Moreover, there are considerable discrepancies between $\alpha_u$ and $\overline \alpha_u$ for the cases for which the latter is defined.  

\begin{table}
\caption{Values of the nonlinear diffusion coefficient obtained via matched asymptotics {\normalfont[}$\alpha_u$ in \cref{alphaV}{\normalfont]} or via closure {\normalfont[}$\overline \alpha_u$ in \cref{alphaV_close}{\normalfont]} for various types of interaction potential in 2 or 3 dimensions. 
}
\label{table:comparison}
\centering
\renewcommand{\arraystretch}{1.3}
\begin{tabular}{@{}lcc|cc|cc@{}}\toprule
& \multicolumn{2}{c}{$\bo{d=1}$}&  \multicolumn{2}{c}{$\bo{d=2}$} &  \multicolumn{2}{c}{$\bo{d=3}$} \\
\cmidrule{2-7} 
 & $\alpha_u$ & $\overline \alpha_u$  & $\alpha_u$ & $\overline \alpha_u$  & $\alpha_u$ & $\overline \alpha_u$   \\ \midrule
$u_\text{HS} = \left(\epsilon/r\right)^\infty$ &  2& $\nexists$ & $\pi$& $\nexists$ & $1.33\pi$ & $\nexists$ \\
$u_\text{SS} = \left(\epsilon/r\right)^4$ & $2.45$ &  $\nexists$  & $1.77\pi$ & $\nexists$ & $4.83\pi$ & $\nexists$\\
$u_\text{LJ} = \left( \frac{\epsilon}{r} \right)^{12} \!- \! \left( \frac{\epsilon}{r} \right)^{6}$ & 1.63  & $\nexists$ & $0.555\pi$ & $\nexists$  & $0.154\pi$ & $\nexists$ \\
$u_\text{EX} = e^{-r/\epsilon}$  & $1.59$ &  $2$  & $1.78\pi$ & $2\pi$ & $7.54\pi$ & $8\pi$\\
$u_\text{YU} = \frac{\epsilon}{r}e^{-r/\epsilon}$ &  $1.80$& $\nexists$& $1.25\pi$& $2\pi$ & $3.47\pi$ & $4\pi$ \\ 
\bottomrule
\end{tabular}
\end{table}

The coefficient  $\overline \alpha_u$ is undefined for inverse-power potentials such as SS and LJ, since the integral in \cref{alphaV_close} is either singular at the origin or at infinity for all possible powers. Therefore, the LMFA is not valid for these potentials. The MAE coefficient $\alpha_u$ exists for the SS potential in \cref{pairpotentialSS} for $\nu>d$, but is undefined for $\nu \le d$. Therefore, we find that MAE is not valid for the Coulomb interaction [$\nu = 1$ in \cref{pairpotentialSS}]. The interpretation is that the Coulomb interaction does not decay sufficiently quickly at infinity and hence the inner region spans the whole configuration space (so there is no outer region where the integral ${\bf G} ({\bf x}_1)$ is negligible as assumed in the MAE derivation). 

\cref{fig:alpha_vs_soft} shows the variation of $\alpha_\text{SS}$ against the hardness parameter $\nu$. As expected, $\lim_{\nu \to \infty} \alpha_\text{SS} = \alpha_\text{HS}$ since the HS potential is the limiting case of the SS potential for $\nu \to \infty$. Also note that the strength of the nonlinear diffusion term, parametrized by $\alpha_\text{SS}$, decreases as the hardness $\nu$ increases. The effect of the softness parameter $1/\nu$ on the diffusion and other transport coefficients was studied in \cite{Heyes:2005ff}; they found that for $\nu \ge 72$ the soft sphere system behaved essentially as a hard sphere system in molecular-dynamics simulations. We find that the relative error between $\alpha_\text{SS}$ and $\alpha_\text{HS}$  for $\nu = 72$ in three dimensions is 2.6\%. 
\def \scc {0.7}
\def \scl {1.0}
\begin{figure}[thb]
\unitlength=1cm
\begin{center}
\psfragscanon
\psfrag{nu}[][][\scl]{$\nu$} \psfrag{a}[][][\scl]{$\displaystyle \frac{\alpha_\text{SS}}{\pi}$}  \psfrag{p1}[][][\scl]{$(a)$}  \psfrag{p2}[][][\scl]{$(b)$} 
\includegraphics[width=.8\textwidth]{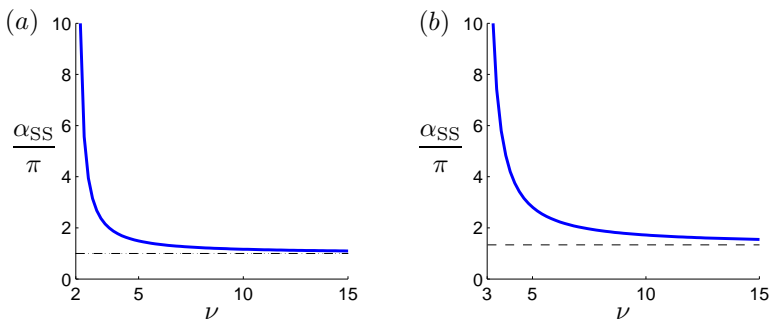}
\caption{Coefficient $\alpha_\text{SS}$ given by \cref{alphaV} for the SS potential \cref{pairpotentialSS}, versus the hardness parameter $\nu$. (a) Two dimensions. (b) Three dimensions. The dash lines show the corresponding $\alpha_\text{HS}$, equal to $\pi$ for $d=2$ and $4\pi/3$ for $d=3$.}
\label{fig:alpha_vs_soft}
\end{center}
\end{figure}

The Morse potential \cref{pair_MO} has well-defined coefficients $\alpha_\text{MO}$ and $\overline \alpha_\text{MO}$  for $C$ and $l$ positive. However, they differ substantially depending on the relative strength $C$ and relative lengthscale $l$ of the repulsion to the attraction, see \cref{fig:alpha_morse}.  We note that both coefficients  $\alpha_\text{MO}$ and $\overline \alpha_\text{MO}$ may be negative for some parameter values, which implies, from \cref{sfpNfinal_eq}, that the nonlinear component of the diffusion coefficient becomes negative. When this occurs, the system enters a so-called catastrophic regime and the particles collapse to a point as $N\to \infty$ \cite{DOrsogna:2006ci}.   
\def \scc {0.7}
\def \scl {1.0}
\begin{figure}[thb]
\unitlength=1cm
\begin{center}
\psfragscanon
\psfrag{c}[][][\scl]{$C$} \psfrag{l}[][][\scl]{$l$}  \psfrag{a}[b][l][\scl]{$\alpha_\text{MO}$ \ (a) }  \psfrag{b}[b][][\scl]{$\alpha_\text{MO}$ \  (b) } 
\includegraphics[height=.29\textwidth]{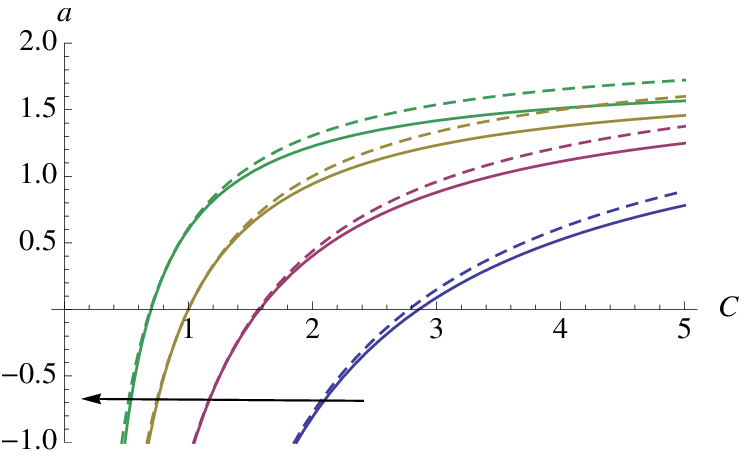} \qquad 
\includegraphics[height=.29\textwidth]{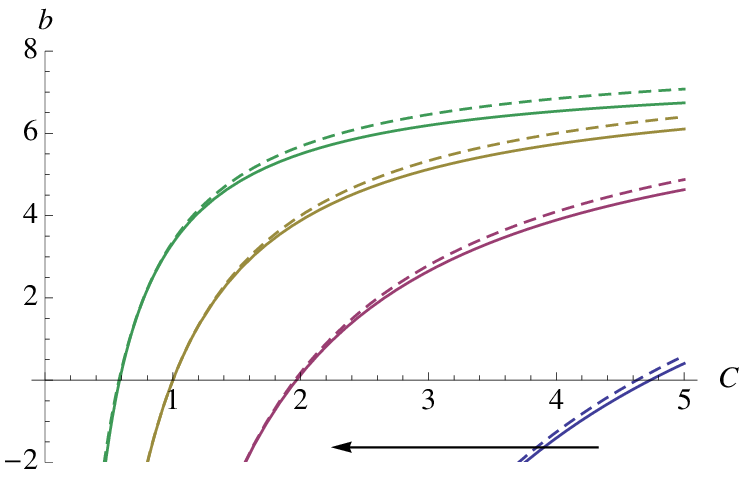}
\caption{Coefficients $\alpha_\text{MO}$ (solid lines) and  $\overline \alpha_\text{MO}$ (dash lines) versus $C$ for 4 values of $l$, $l=0.6, 0.8, 1, 1.2$ (arrow indicating the direction of increasing $l$). (a) Two dimensions. (b) Three dimensions. The system is said to be in a catastrophic regime when $\overline \alpha_u<0$.}
\label{fig:alpha_morse}
\end{center}
\end{figure}

Evaluation of $\alpha_u$ and $\overline \alpha_u$ provides a straightforward way to determine the type of dynamics the interacting particle system has. Initially it is not so easy to discern whether the potential is ``short-range enough'' and, as a result, which method is more suitable to obtain a reduced population-level model. The nonlinear diffusion coefficient \cref{sfpNfinal_eq}, $\alpha_u (N-1) \epsilon^d p$, gives an idea of the strength of the interaction and whether the MAE will be the appropriate method (in particular, one can compute the effective volume fraction as described in \cref{sec:softhard_pot} and ascertain whether the low-volume fraction assumption holds). Regarding the MFA closure, one should keep in mind that $\overline \alpha_u$ not being defined does not necessarily mean that standard MFA fails, as mentioned in \cref{sec:soft_FP}; it could be that the short-range assumption used to obtain the nonlinear diffusion equation \cref{sfpNfinal_close} does not hold and that, instead, one should keep the original MFA integro-differential model \cref{mean_field}. 
Conversely, if $\overline \alpha_u$ is not defined due to the singular behavior of the potential at the origin (which is the case for all those seen in \cref{table:comparison}), this is an indication that the convolution in the original MFA will not exist.

\subsection{Comparison with the particle-level model}
\label{sec:numerics}

In this section we compare the macroscopic models obtained via the closures MFA and KSA, and via the MAE method we have introduced to each other as well  as to numerical simulations of the stochastic particle system.

We use the open-source C\texttt{++} library Aboria \cite{aboria,Robinson:2017vxa} to perform the particle-level simulations. The overdamped Langevin equation \cref{ssde} is integrated using the Euler-Maruyama method and a constant timestep $\Delta t$, leading to an explicit update step for each particle given by
\begin{equation}\label{eq:particle}
\mathbf{X}^{m+1}_i = \mathbf{X}^m_i + \sqrt{2D\Delta t}\Delta W^m + {\bf f}  ( {\bf X}_i^m ) \Delta t  - \sum_{j \ne i}  \nabla_{{\bf x}_i} u(\|{\bf X}_i^m - {\bf X}_j^m\|) \Delta t,
\end{equation}
where $\Delta W^m$ is a $d$-dimensional normally distributed random variable with zero mean and unit variance. We choose the timestep $\Delta t$ so that the results are converged (that is, there is no change in the results for smaller timesteps). For all simulations in this paper a timestep of $\Delta t = (0.1\epsilon)^2/2D$ was sufficient for convergence, leading to an average diffusion step size of $0.1\epsilon$.

A naive implementation of the particle force interaction term over all particle pairs would lead to a large number ($N^2$) of potential pair interactions to perform. To improve the efficiency of the code, we take advantage of the compact nature of the potentials and restrict particle interaction to those pairs that are within a certain cutoff length $c < 6\epsilon$. All particle pairs $i,j$ separated by a distance greater than this cutoff will implicitly be given a pair potential of $u(\|{\bf X}_i - {\bf X}_j\|) = 0$.

In order to compare the particle-level models with the PDE models, we perform $R$ independent realizations and output the positions of all $NR$ particles at a set of output time points. A histogram of the positions is calculated and then scaled to produce a discretized density function that can be compared with the PDE models. To generate the two-particle density function, we create a two-dimensional histogram of the positions of each particle pair $(i,j)$ and scale it accordingly to produce a two-particle density.

First, we consider a one-dimensional problem in $\Omega = [0, 1]$ with periodic boundary conditions. We compare estimates of the one-particle density $p(x_1,t)$ as well as the two-particle density $P_2(x_1, x_2,t)$ obtained from simulating \cref{eq:particle} to solutions of the same quantities using the KSA, MFA or MAE models. We start with a set of parameters in which the potential is sufficiently long range that the system is not really in the low-density limit with the number of particles we use, and then consider an example with a strongly repulsive short-ranged potential. In all the examples we set the external drift to zero.

We begin by presenting the models for the three approaches for $d=1$. The MFA reads, combining \cref{sfp_int} and \cref{Bcclosed}, 
\begin{equation} 
\label{MFA}
\frac{\partial p}{\partial t}  = \frac{\partial}{\partial x_1}  \left[ \frac{\partial p }{\partial x_1}  + (N-1 ) \, p\! \int_{\Omega} p( x_2, t) f_u(x_1, x_2)   \,  \ud x_2  \right],
 \end{equation}
where $p = p(x_1,t) $ unless explicitly written and  $f_u(x_1, x_2) := \frac{\partial}{\partial x_1}  u(|x_1 - x_2|)$. To solve \cref{MFA} we use a second-order accurate finite-difference approximation with $M$ grid points in space and the method of lines in time with the inbuilt Matlab ode solver \texttt{ode15s}. To evaluate the integral in \cref{MFA}, we use the periodic trapezoid rule (which converges exponentially fast for smooth integrands \cite[Chapter 12]{Trefethen:2000jp}). To avoid evaluating the interaction potential at zero, we shift the grid for $x_2$ by $h/2$, where $h = 1/M$ is the grid spacing. The density $p(x_2,t)$ is approximated from $p(x_1,t)$ using $p(x_2,t) \approx (p(x_2-h/2,t) + p(x_2+h/2,t))/2$. Because of the convolution term, the discretization matrix of \cref{MFA} is full, making the numerical solution of the MFA computationally expensive. An alternative would be to use a nonuniform mesh with more points near the diagonal or an adaptive grid scheme such as that presented by Carrillo and Moll \cite{Carrillo:2009dz}, which uses a transport map between the uniform density and the unknown density $p$ such that more grid points are placed where the density is higher. 

The KSA closure model is the coupled system for $p(x_1,t)$ and $P_2(x_1, x_2,t)$, given by
\begin{subequations} \label{KSA}
\begin{equation} 
\label{KSA_p}
\frac{\partial p}{\partial t} = \frac{\partial}{\partial x_1}  \left[ \frac{\partial p }{\partial x_1} + (N-1 ) \! \int_{\Omega} P_2  f_u(x_1, x_2)  \,  \ud x_2  \right],
 \end{equation}
\begin{align} \label{KSA_P}
\begin{aligned} 
\frac{\partial P_2}{\partial t}
=    & \frac{\partial}{\partial x_1}  \! \bigg[ \frac{\partial P_2}{\partial x_1}   +  f_u(x_1, x_2) P_2  \\ & \hspace{2cm} +  (N-2) \tfrac{ P_2 (x_1,x_2,t) } {p (x_1,t) p (x_2,t)} \! \int_\Omega  \tfrac{ P_2 (x_1,x_3,t) P_2 (x_2,x_3,t) } {p (x_3,t) }  f_u(x_1, x_3)  \, \ud x_3
  \bigg]   \\
  & +  \frac{\partial}{\partial x_2}  \! \bigg[ \frac{\partial P_2}{\partial x_2}   +  f_u(x_2, x_1) P_2  \\ & \hspace{2cm}  +  (N-2) \tfrac{ P_2 (x_1,x_2,t) } {p (x_1,t) p (x_2,t)} \! \int_\Omega  \tfrac{ P_2 (x_1,x_3,t) P_2 (x_2,x_3,t) } {p (x_3,t) }  f_u(x_2, x_3)  \, \ud x_3
  \bigg], 
\end{aligned}
\end{align}
\end{subequations}
where $p = p(x_1,t) $ and $P_2 = P_2(x_1,x_2, t)$ unless explicitly written. We solve in the computational spatial domain $\Omega^2$ and therefore the KSA model requires a two-dimensional grid with full matrices, making it very computationally expensive. As with the MFA, we shift the grid for $x_2$ by $h/2$ to avoid any issues with $f_u(x_1, x_2)$ when the interaction potential $u$ is singular at the origin. In the first integral of \cref{KSA_P}, the coordinate $x_3$ is evaluated on the same grid as $x_2$, and in the second integral it is evaluated on the same grid as $x_1$. 
An alternative implementation of the KSA system is to solve for $P_2$ only and evaluate the one-particle density as $p(x_1,t) = \int_\Omega P_2(x_1,x_2,t) \, \ud x_2$ (replacing \cref{KSA_p}) \cite{Middleton:2014fa}. While this is true in the infinite BBGKY hierarchy, we note that once the KSA closure \cref{kirkwood} is adopted, the resulting model is not necessarily equal to the KSA system \cref{KSA}.

Finally, the reduced model from the MAE method reads
\begin{equation} 
\label{MAE}
\frac{\partial p}{\partial t} = \frac{\partial}{\partial x_1}  \left \{ \left [1+ \alpha_u (N-1) \epsilon^d p \right] \frac{\partial p }{\partial x_1} \right\} ,
 \end{equation}
where $\alpha_u = 2 \int_0^\infty  (1-e^{-u (\epsilon r)} ) \, \ud r$. Noting that the right-hand side in \cref{MAE} can be written as $[p + \frac{1}{2} \alpha_u (N-1) \epsilon^d p^2]_{x x}$, the numerical implementation is straightforward and, since the discretization matrix is banded, very efficient.

In the first example, we use the exponential potential $u_\text{EX}$ \cref{pairpotentialEX} with $\epsilon = 0.05$ in a system in $N=16$ particles. The particles are initially distributed according to $p(0) = 0.5[\tanh(\beta(x-\theta))+\tanh(\beta(1-\theta-x))]$, with $\beta = 30$ and $\theta = 0.2$, and we let them diffuse until time $T_f = 0.02$. The evolution in time of the one-particle density obtained with the various methods is shown in \cref{fig:1dcase1}. We observe very good agreement between the stochastic simulations and the MFA and KSA models, whereas the MAE model slightly underestimates the diffusion strength. This can be explained since, for the chosen values of $\epsilon$ and $N$, the potential is not short-ranged and the assumptions of the MAE model break down. In particular, using \cref{epeff} we find that the effective hard-sphere radius is $\epsilon_\text{eff} = 0.04$ and hence the effective volume fraction is 0.64. Similarly, the solution obtained via the LMFA (shown at $T_f$ in \cref{fig:1dcase1}(b)) differs noticeably to the MFA solution. 
\def \scc {0.7}
\def \scl {1.0}
\begin{figure}[bh!]
\unitlength=1cm
\begin{center}
\psfrag{a}[][][\scl]{(a)} \psfrag{b}[][][\scl]{(b)}
\psfrag{x}[][][\scl]{$x_1$} \psfrag{t}[][][\scl]{$t$} \psfrag{q}[r][][\scl][-90]{$p$}
\includegraphics[width=.495\textwidth]{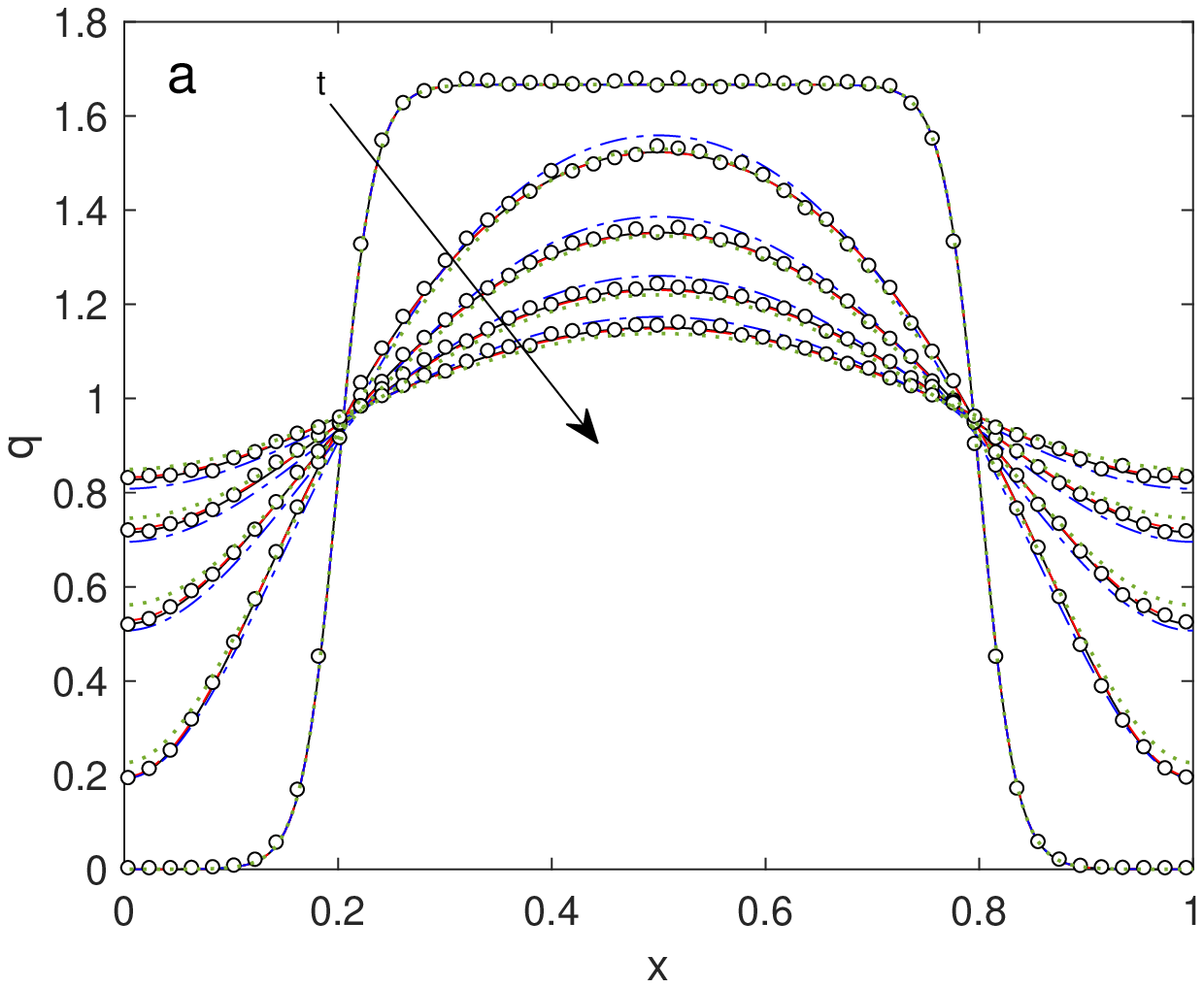} \hfill \includegraphics[width=.495\textwidth]{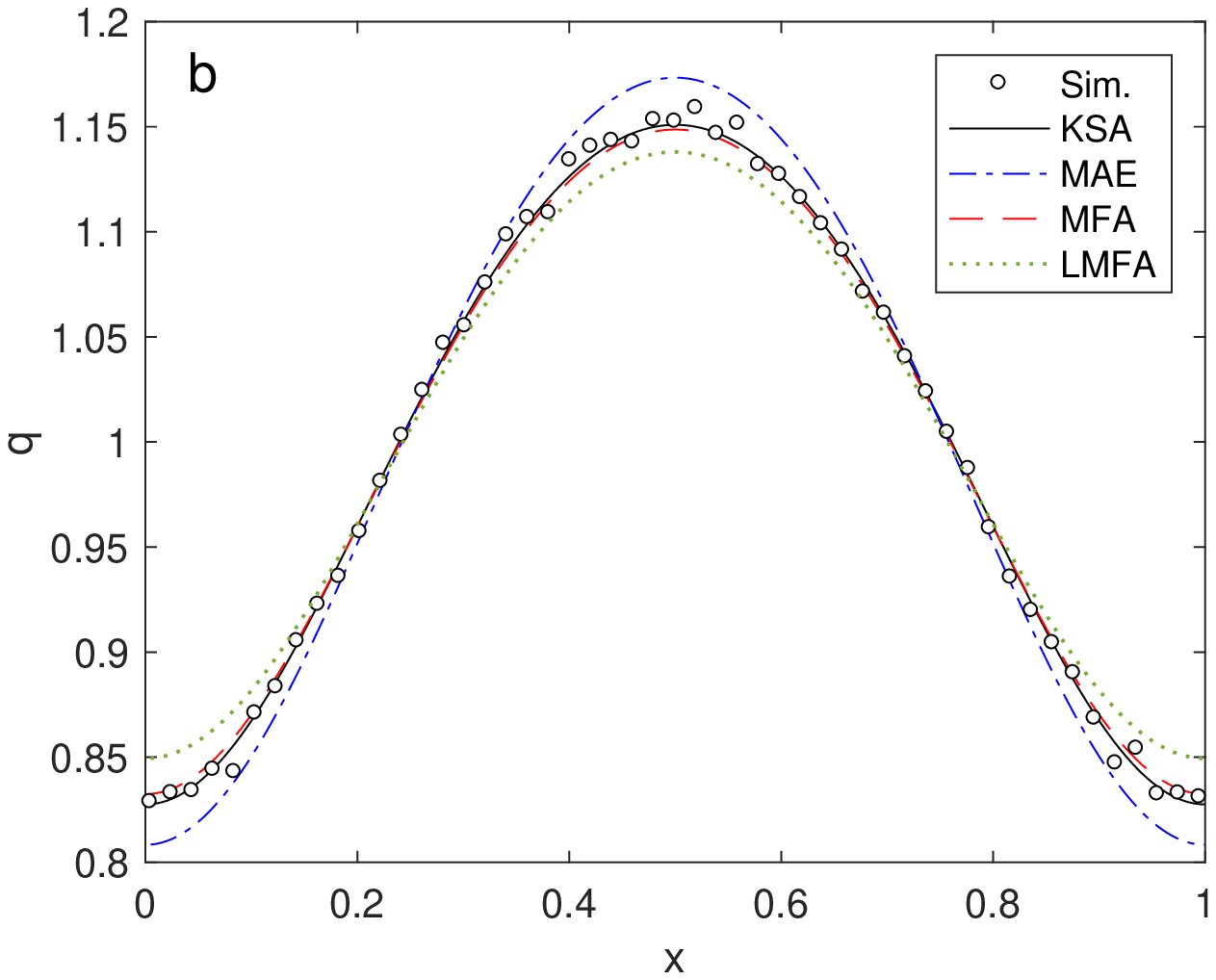}
\caption{One-particle density function $p(x_1,t)$ at times (a) $t = 0, T_f/4, T_f/2, 3T_f/4, T_f$ and (b) $t = T_f$  with $T_f = 0.02$, interaction potential $u_\text{EX}(r) = \exp(-r/\epsilon)$, $\epsilon = 0.05$, and $N= 16$. The initial data is $p(x_1,0) = 0.5[\tanh(\beta(x_1-\theta))+\tanh(\beta(1-\theta-x_1))]$, with $\beta = 30$ and $\theta = 0.2$ and we use periodic boundary conditions. Solution of the KSA model \cref{KSA}, the MAE model \cref{MAE}, the MFA model \cref{MFA}, the LMFA model \cref{sfpNfinal_close}, and histogram computed from $R=3\times10^5$ realizations of the particle-level simulation \cref{eq:particle}. We use $M=200$ grid points in each direction to solve the KSA, MAE and MFA models.} 
\label{fig:1dcase1}
\end{center}
\end{figure}

\begin{figure}[tbh!]
\unitlength=1cm
\begin{center}
\includegraphics[width=\textwidth]{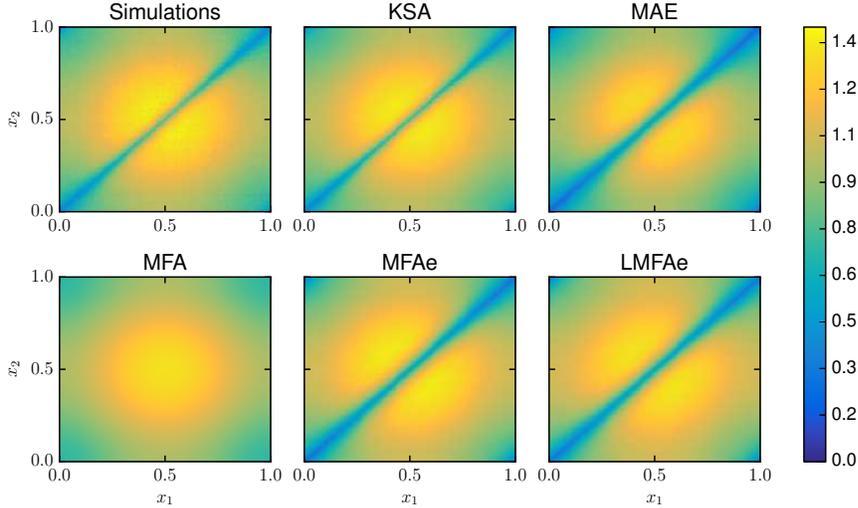}
\caption{Two-particle density function $P_2(x_1,x_2,t)$ at time $T_f = 0.02$ for the parameters in \cref{fig:1dcase1}. The initial data is $P_2(x_1, x_2, 0) = p(x_1,0) p(x_2,0)$. The plots MFAe and LMFAe use the approximation $P_2(x_1, x_2, t) \approx p(x_1, t) p(x_2,t) \exp(-u(|x_1-x_2|))/\iint \exp(-u(|x_1-x_2|)) \ud x_1 \ud x_2$, where $p$ is computed with the MFA or LMFA model, respectively.} 
\label{fig:2dcase1}
\end{center}
\end{figure}

\Cref{fig:2dcase1} shows the corresponding two-particle density function at the final time $T_f$. For the KSA, $P_2$ is solved for as discussed above. For the MFA and MAE models, $P_2$ is given by \cref{closure_approx} and \cref{MAE_p2} respectively. As before, $x_2$ is shifted by half the grid size and $p(x_2,t)$ is approximated from $p(x_1,t)$ using the centered average. 
The correlation between particles can be seen by the drop in probability at the diagonal $x_1 = x_2$ in the simulations as well as in the KSA and MAE plots. Conversely, the MFA  misses the correlation between particles, as expected from the ansatz $P_2(x_1, x_2, t) = p(x_1, t) p(x_2,t)$. The differences in $P_2$ between the models can be clearly seen in \cref{fig:case12centrelines}(a), which shows a plot of $P_2(x_1,x_2,T_f)$ along $x_2 = 0.5$. 
\def \scc {0.7}
\def \scl {1.0}

Next we consider an example using a more repulsive interaction potential. In particular, we consider a smoothed version of the Yukawa potential \cref{pairpotentialYU}, namely $u (r) =(\epsilon /\sqrt{r^2 + \delta^2}) \exp^{-r/\epsilon}$ with $\epsilon = 0.01$ and $\delta = 0.002$. We choose to smooth the potential so that particles can still swap positions and so that we can use the closure models KSA and MFA (in one dimension the singularity at zero poses problems in the convolution terms, see \cref{table:comparison}). We run a simulation with $N=20$ particles up to time $T_f = 0.02$. For these parameters, the effective hard-sphere radius is $\epsilon_\text{eff} = 0.009$ and the effective volume fraction is 0.18. Thus we expect the MAE model to perform better than in the previous example where the potential was not short range. The one-particle density plots are shown in \cref{fig:1dcase2}, and the results for the two-particle density function are shown in \cref{fig:2dcase2}. We see that the MFA closure solution diffuses substantially faster than the KSA closure solution, whereas the MAE model spreads slightly slower than the KSA one (\cref{fig:1dcase2}). The particle-level simulations match the KSA solution closely over all times. The two-particle density functions for the particle simulation, the KSA and the MAE solution are indistinguishable at time $T_f = 0.02$, whereas the MFA misses the drop in probability for closely spaced particles, as expected due to its assumption of no particle correlations (\cref{fig:2dcase2,fig:case12centrelines}(b)).
\def \scc {0.7}
\def \scl {1.0}
\begin{figure}[tbh]
\unitlength=1cm
\begin{center}
\psfragscanon
\psfrag{a}[][][\scl]{(a)} \psfrag{b}[][][\scl]{(b)}
\psfrag{x}[][][\scl]{$x_1$} \psfrag{t}[][][\scl]{$t$} \psfrag{q}[r][][\scl][-90]{$p$}
\includegraphics[width=.495\textwidth]{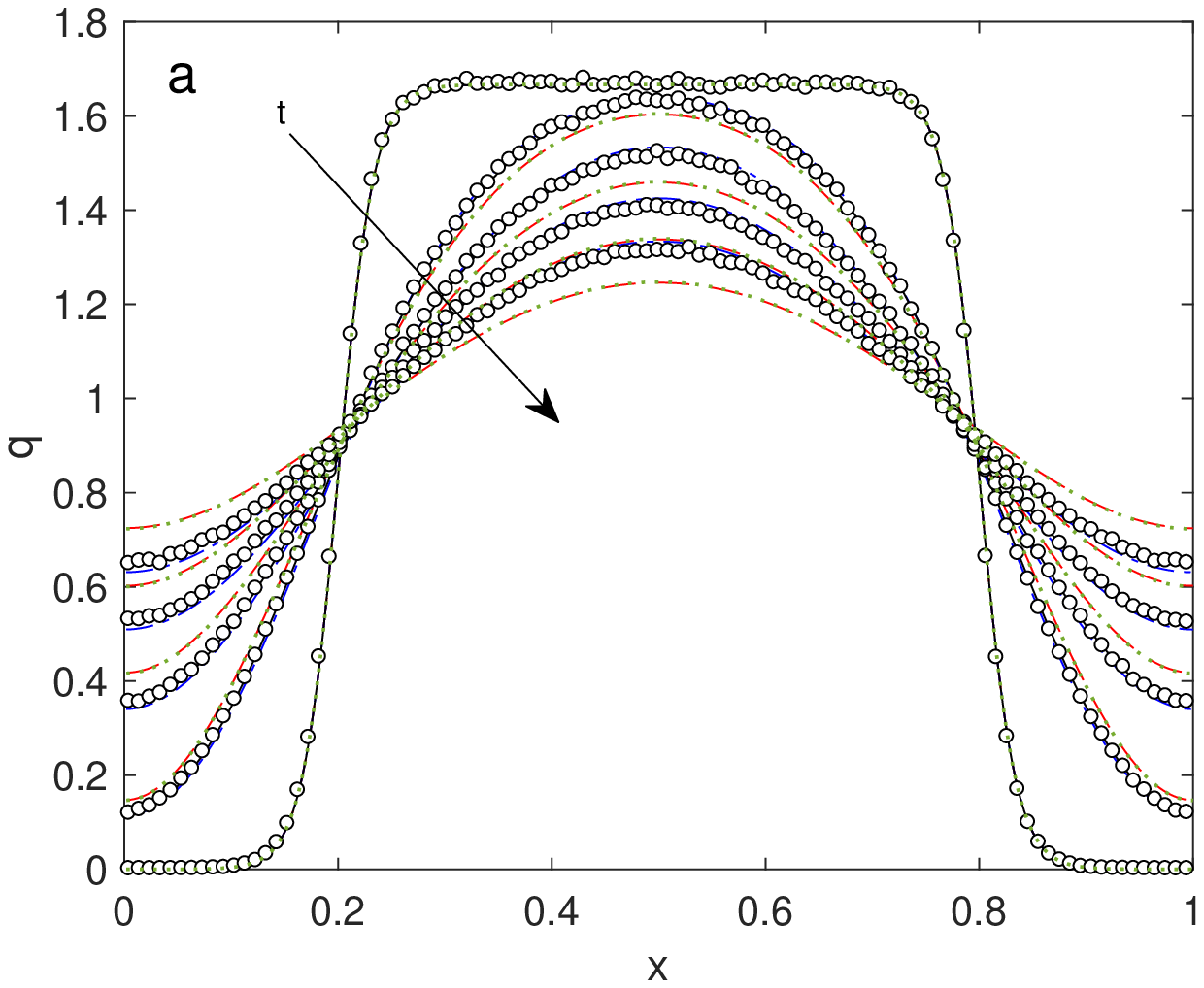} \hfill \includegraphics[width=.495\textwidth]{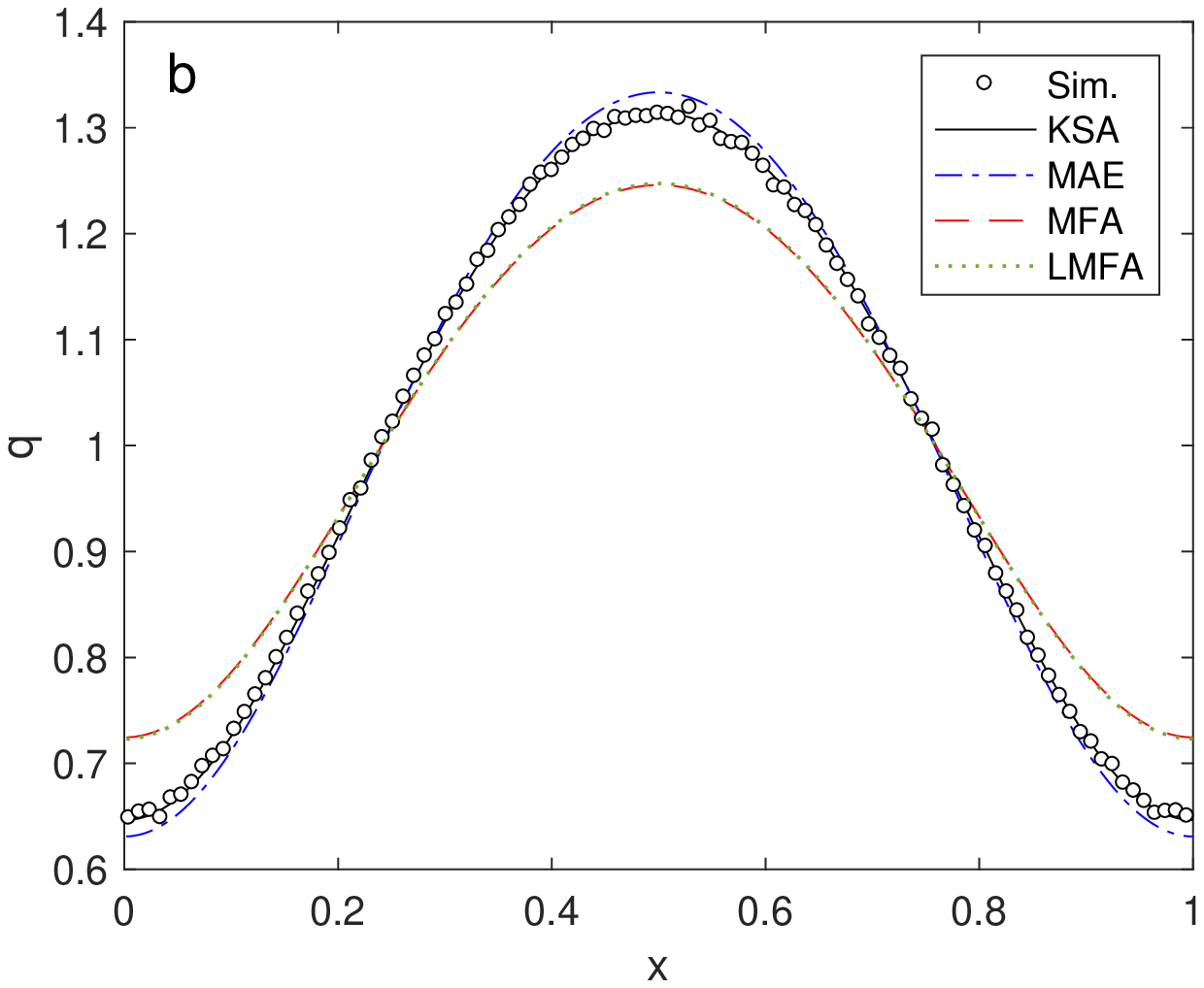}
\end{center}
\caption{One-particle density function $p(x_1,t)$ at times (a) $t = 0, T_f/4, T_f/2, 3T_f/4, T_f$ and (b) $t = T_f$ with $T_f = 0.02$, interaction potential $u (r) = (\epsilon/\sqrt{r^2 + \delta^2}) \exp^{-r/\epsilon}$, $\epsilon = 0.01$, $\delta = 0.002$, and $N= 20$. The initial data is $p(x_1,0) = 0.5[\tanh(\beta(x_1-\theta))+\tanh(\beta(1-\theta-x_1))]$, with $\beta = 30$ and $\theta = 0.2$ and we use periodic boundary conditions. Solution of the KSA model \cref{KSA}, the MAE model \cref{MAE}, the MFA model \cref{MFA}, the LMFA model \cref{sfpNfinal_close}, and histogram computed from $R=3\times10^5$ realizations of the particle-level simulation \cref{eq:particle}. We use $M=200$ grid points in each direction to solve the KSA, the MFA and the MAE models.
} 
\label{fig:1dcase2}
\end{figure}

\def \scc {0.7}
\def \scl {1.0}
\begin{figure}[tbh!]
\unitlength=1cm
\begin{center}
\includegraphics[width=\textwidth]{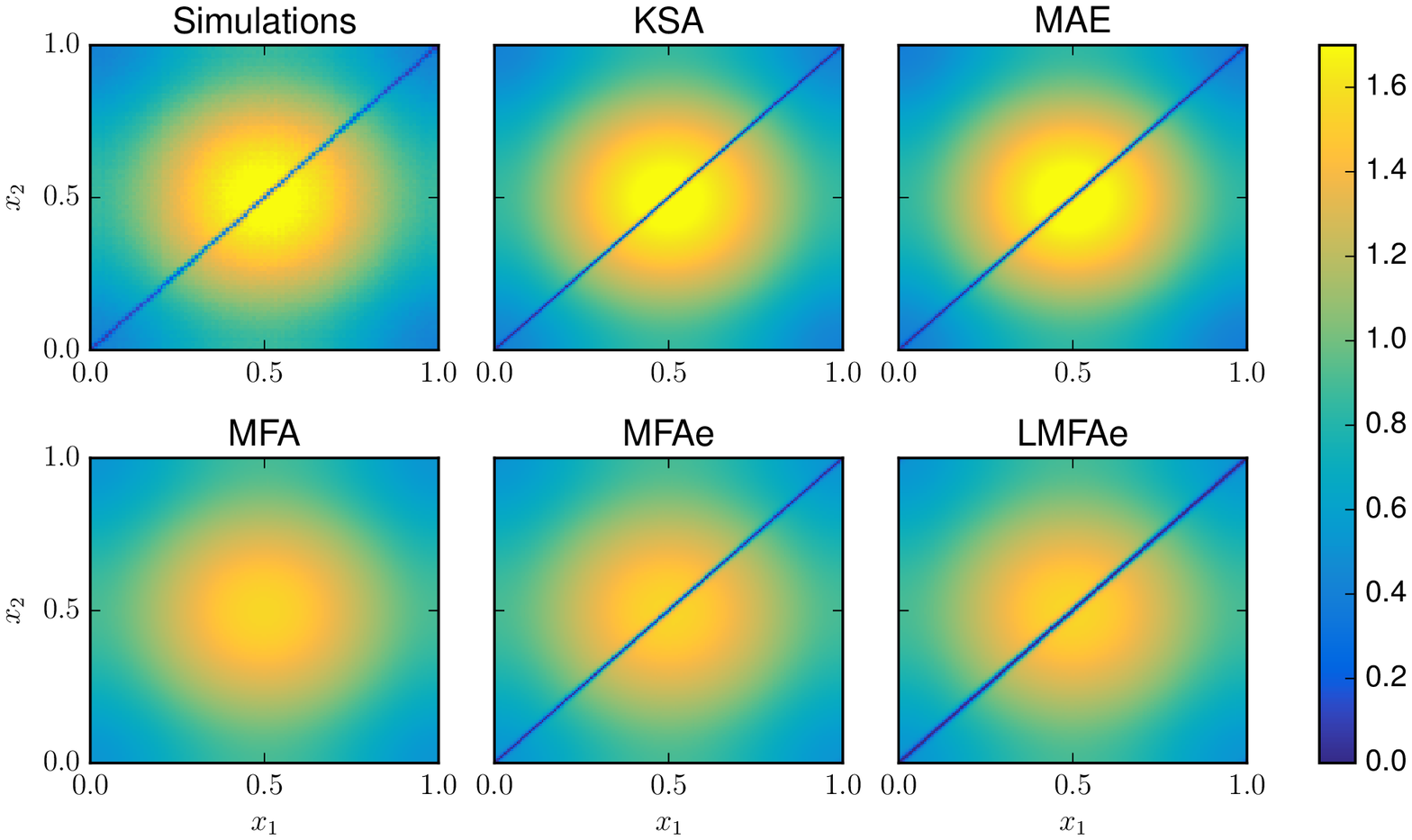}
\caption{Two-particle density function $P_2(x_1,x_2,t)$ at time $T_f = 0.02$ for the parameters in \cref{fig:1dcase2}. The initial data is $P_2(x_1, x_2, 0) = p(x_1,0) p(x_2,0)$. The plots MFAe and LMFAe use the approximation $P_2(x_1, x_2, t) \approx p(x_1, t) p(x_2,t) \exp(-u(|x_1-x_2|))/\iint \exp(-u(|x_1-x_2|)) \ud x_1 \ud x_2$, where $p$ is computed with the MFA or LMFA model, respectively.} 
\label{fig:2dcase2}
\end{center}
\end{figure}

\begin{figure}[tbh]
\unitlength=1cm
\begin{center}
\psfragscanon
\psfrag{a}[][][\scl]{(a)} \psfrag{b}[][][\scl]{(b)}
\psfrag{x}[][][\scl]{$x_1$} \psfrag{t}[][][\scl]{$t$} \psfrag{P}[r][][\scl][-90]{$P_2$}
\includegraphics[width=.495\textwidth]{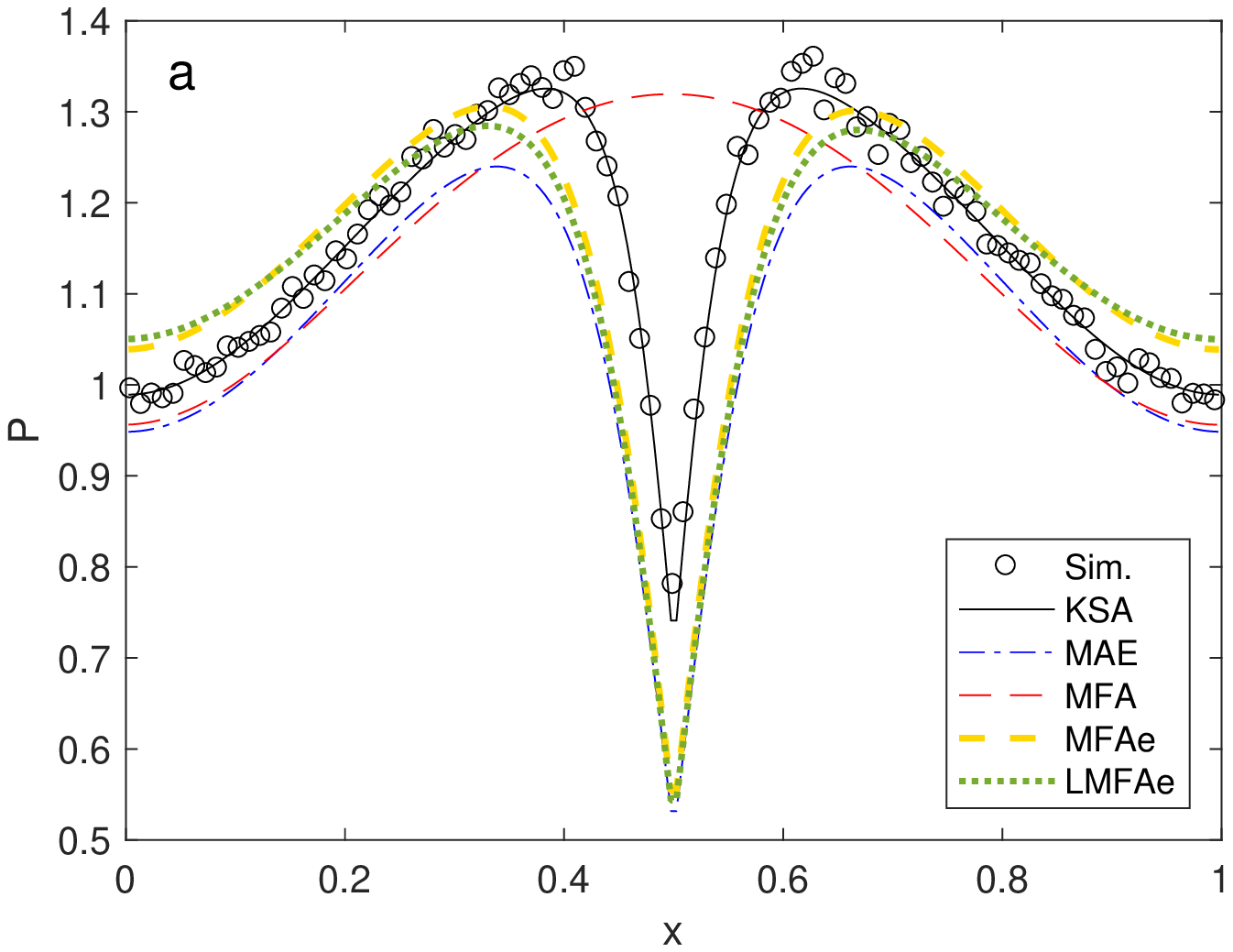} \hfill \includegraphics[width=.495\textwidth]{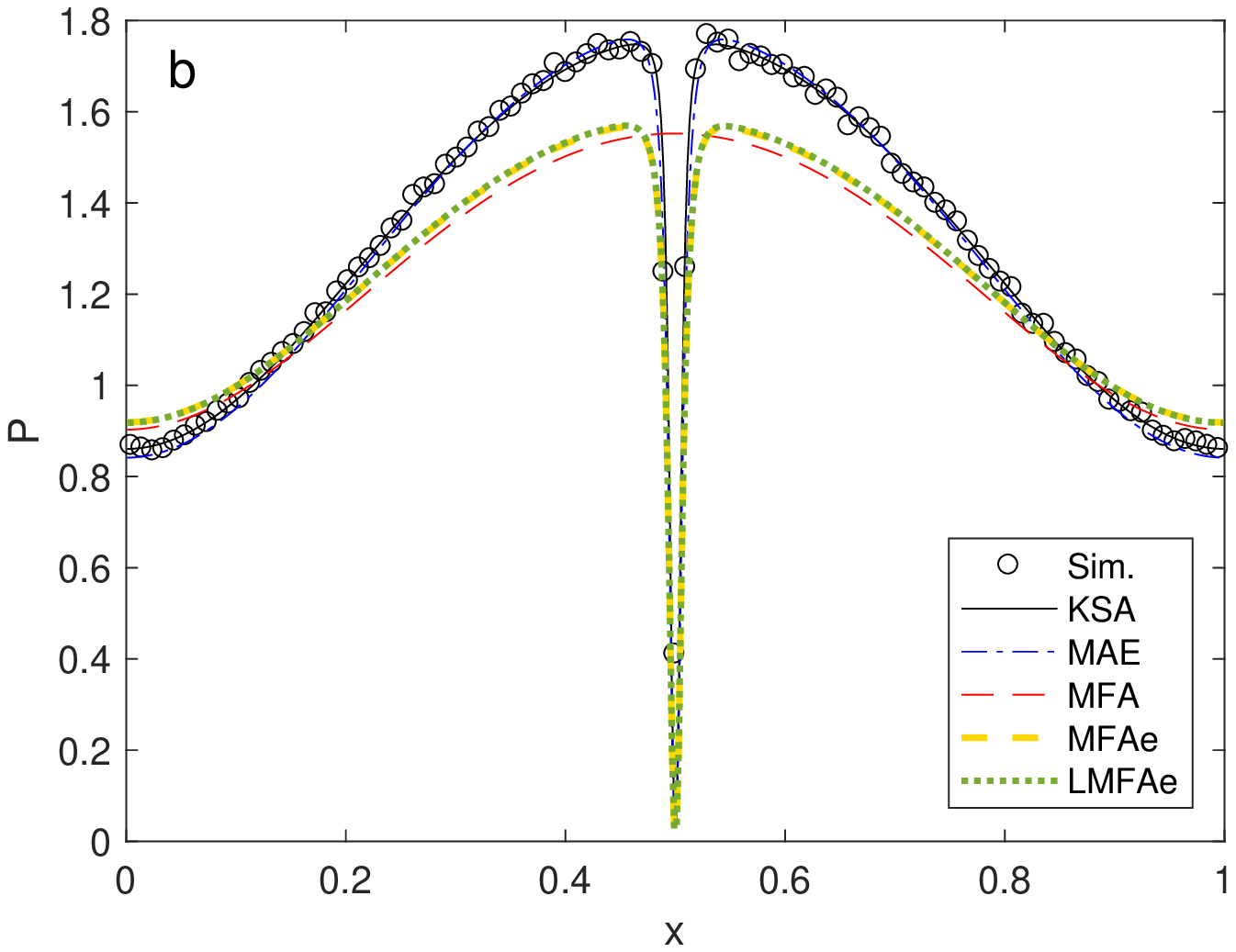}
\end{center}
\caption{Two-particle density function $P_2(x_1,x_2,t)$ along the line $x_2 = 0.5$ at time $T_f = 0.02$ corresponding to the two-dimensional plots shown in (a) \cref{fig:2dcase1} and (b) \cref{fig:2dcase2}.
} 
\label{fig:case12centrelines}
\end{figure}

Generally, we see that MAE is good for strongly repulsive interactions (such as in the example in \cref{fig:1dcase2,fig:2dcase2}) while MFA is better for softer or longer-range interactions (\cref{fig:1dcase1,fig:2dcase1}). The KSA provides a good approximation in all cases. When the two-particle density is required in a system with long-range interactions, our results may one lead to think that one should use the KSA, since the MFA does not capture the correlation in $P_2$. However, we suggest that a slight modification in the MFA can provide a reasonable approximation to $P_2$ also. Specifically, one could still use the standard MFA closure $P_2(x_1, x_2, t) = p(x_1, t) p(x_2,t)$  to compute the one-particle density, but then use  $P_2(x_1, x_2, t) \approx p(x_1, t) p(x_2,t) \exp(-u(|x_1-x_2|)) /C$ as an approximation of the two-particle density, where $C =\iint \exp(-u(|x_1-x_2|)) \ud x_1 \ud x_2$ is a normalization constant. This modification of $P_2$, using either the MFA or the LMFA to compute $p$, is shown as MFAe and LMFAe respectively in \cref{fig:2dcase1,fig:2dcase2,fig:case12centrelines}.  
We see that MFAe and LMFAe provide a better approximation of $P_2$ than MFA and MAE in the first example with longer-range interactions (\cref{fig:case12centrelines}(a)), whereas in the second example (with strong repulsion) the three approximations MFA, MFAe, and LFMAe are poor (\cref{fig:case12centrelines}(b)). 

Finally, we consider a two-dimensional example in $\Omega = [0,1] ^2$. We choose a system of $N =400$ Yukawa-interacting particles, with interaction potential $u_\text{YU}$  \cref{pairpotentialYU} and $\epsilon = 0.01$, initially distributed according to a normal distribution in $x$ with mean $0.5$ and standard deviation $\sigma = 0.05$. The effective hard-sphere radius is $\epsilon_\text{eff} = 0.0112$, giving an effective volume fraction of 0.04. 
Being in two dimensions, solving the KSA model would require solving a system of $M^4$ equations, where $M$ is the number of grid points in one direction. Because of the short-range nature of the pair potential $u_\text{YU}$, we require a large number of points $M$ to resolve the interaction near the origin. As a result, solving the KSA model for this system becomes computationally impractical, and we compare the MFA and the MAE methods only.  We also solve the interaction-free model ($\epsilon=0$) for reference. We plot the comparison at times $T_f/2$ and $T_f$ in  \cref{fig:final_yu1d}. We observe a very good agreement between the stochastic simulations of the particle system and the MAE model, whereas the MFA model overestimates the diffusion strength. Because of the short-range nature of the potential, we find no noticeable difference between the solution of the MFA \eqref{mean_field} and the localized version LMFA \cref{sfpNfinal_close}, which is computationally much easier to solve (the two curves would lie on top of each other in \cref{fig:final_yu1d}, the norm of the relative error is of order $10^{-4}$).
As expected, the interaction-free case spreads the slowest of all models since the nonlinear diffusion term is set to zero.

\def \scc {0.7}
\def \scl {1.0}
\begin{figure}[tbh!]
\unitlength=1cm
\begin{center}
\psfragscanon
\psfrag{x}[][][\scl]{$x_1$} \psfrag{a}[][][\scl]{(a)}  \psfrag{b}[][][\scl]{(b)} 
\psfrag{q}[r][][\scl][-90]{$p$} 
\psfrag{Poten1}[][][\scc]{P1} 
\psfrag{Poten2}[][][\scc]{P2} 
\psfrag{Clousure}[][][\scc]{C}
\psfrag{Points}[][][\scc]{$\epsilon=0$} 
\psfrag{NumP1}[][][\scc]{P1}
\psfrag{NumP2}[][][\scc]{P2}
\includegraphics[width=.48\textwidth]{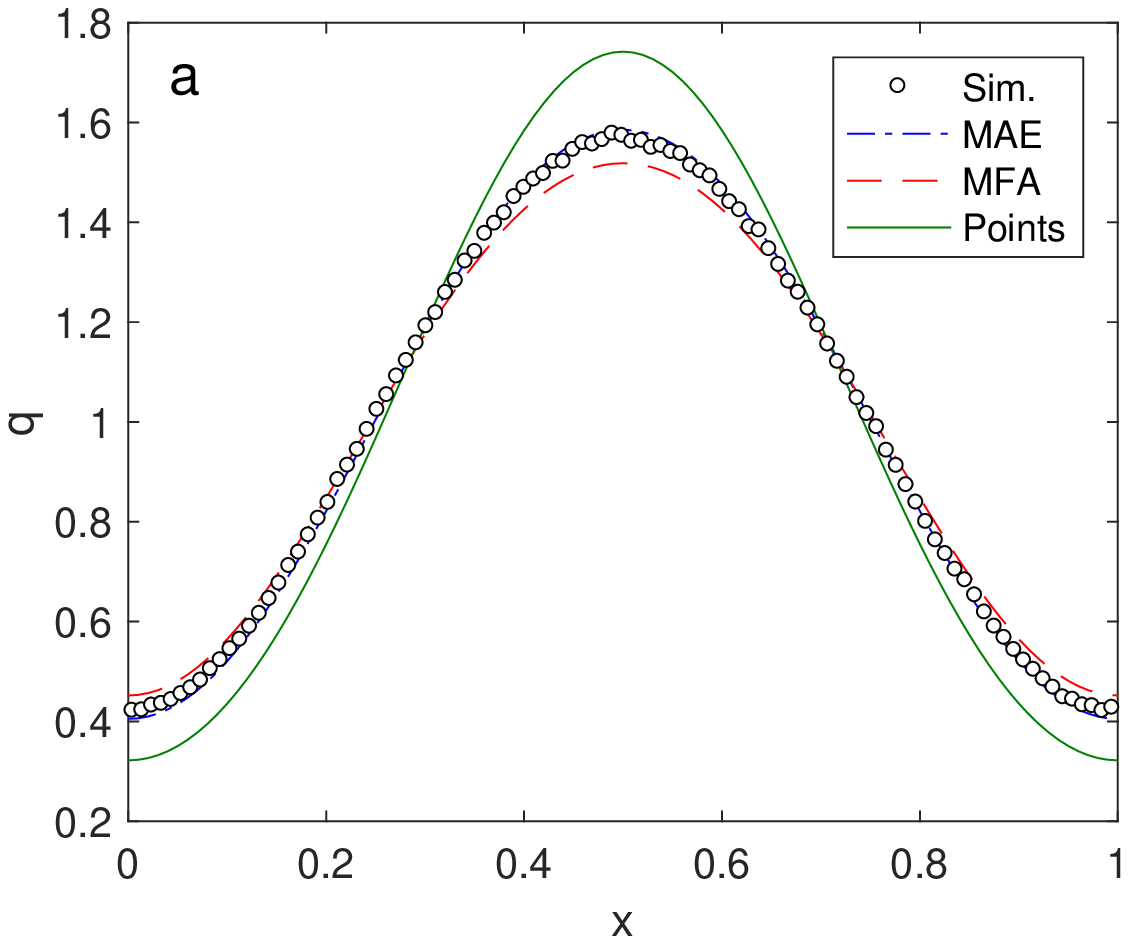} \hfill \includegraphics[width=.48\textwidth]{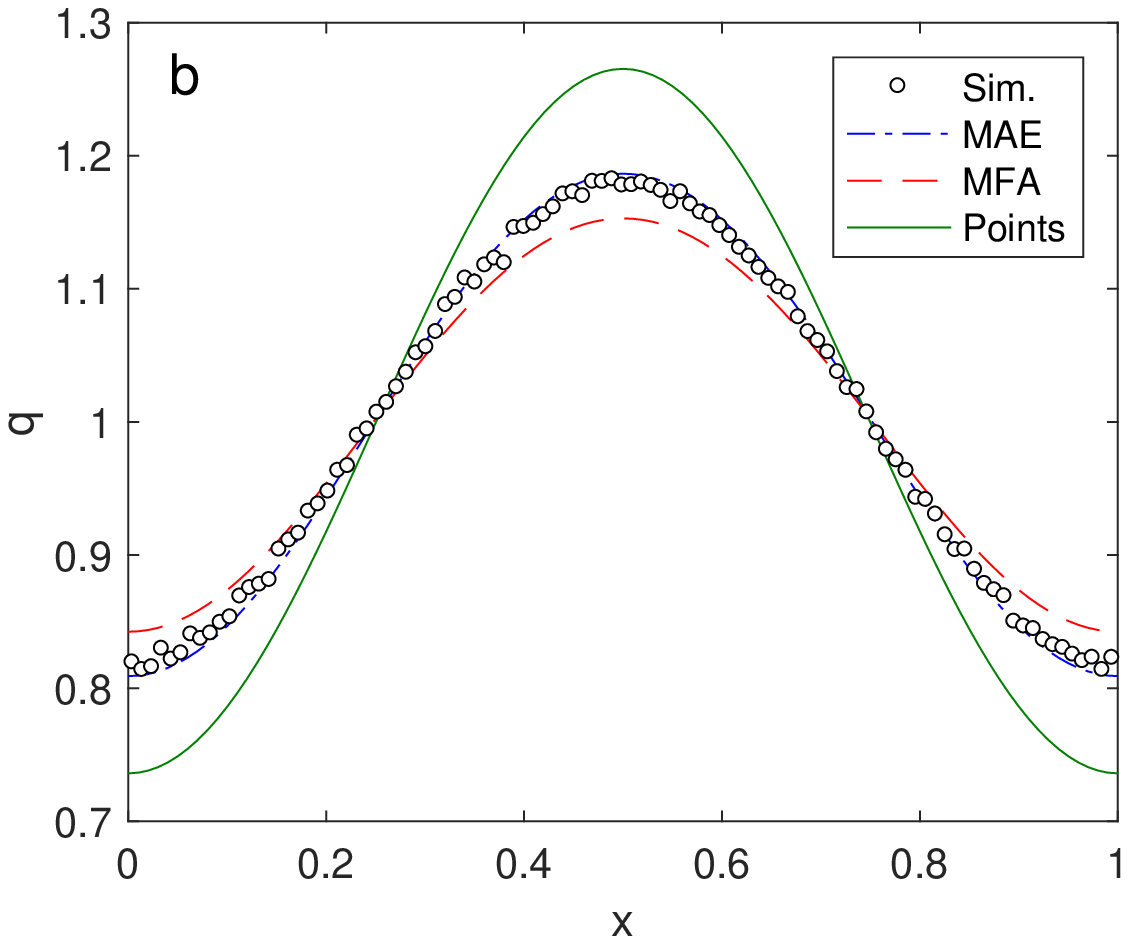}
\caption{
One-particle density function $p(\bfx_1,t)$ at $y_1 = 0$ at times $t= 0.025$ (left) and $t= 0.05$ (right) with interaction potential $u_\text{YU} = (\epsilon/r) \exp(-r/\epsilon)$, $\epsilon=0.01$, and $N=400$. We use 1d-normally distributed initial data ($\mu=0.5$ and $\sigma = 0.05$) and periodic boundary conditions. Solution of the MAE model \cref{sfpNfinal}, the MFA model  \cref{mean_field}, the interaction-free model (setting $\epsilon=0$ in \cref{sfpNfinal}), and histogram computed from $R=1.5\times10^3$ realizations (averaged over the $y_1$-dimension) of the particle-level model \cref{eq:particle}. We use $M=200$ grid points in each direction to solve the MFA and MAE models.}
\label{fig:final_yu1d}
\end{center}
\end{figure}

\section{Discussion and conclusions} \label{sec:conclusions}

We have studied a system of Brownian particles
interacting via a short-range repulsive potential $u$,
and have discussed several ways to obtain a population-level model for
the one-particle density. In particular, we have considered two
common closure approximations  and presented an alternative method
based on matched asymptotic expansions (MAE). The MAE method has the
advantage that it is systematic and works well for very short-ranged
 potentials, especially singular potentials for which common  closure
 approximations can lead to ill-posed models.\footnote{These models are
   sometimes used regardless, with a numerical discretisation
   providing an ad hoc  regularisation of the
   convolution integral \cite{Horng:2012io}.}
The MAE result is a nonlinear
 diffusion equation
similar to our previous work for hard-spheres \cite{Bruna:2012cg}, with 
the  coefficient of the nonlinear term $\alpha_u$ depending on the
potential  through (\ref{alphaV}).

We have performed Monte Carlo simulations of the stochastic particle
system in one and two dimensions, and compared the results with the
solution of the 
MAE model and two common closure approximations: the mean-field
approximation (MFA) and the Kirkwood Superposition Approximation
(KSA). While MFA closes the system at the level of the 
one-particle density $p$, the KSA closes it at the level of the two-particle
density $P_2$. We have tested the models in examples with long-
and short-range interactions.  

We found that the KSA agreed well with the stochastic simulations in
both scenarios, but we could only use it in the one-dimensional
examples due to its high
computational cost. This is because the discretisation of the
convolution term yields a full 
 matrix, making the method impractical to use in two or
three dimensions, especially for strongly repulsive potentials that
require a very fine mesh in the region where two particles are in
close proximity. 
This is also true, but to a lesser extent, for the MFA model, which
captured well the behavior of the system with a long-range potential
but was outperformed by MAE in examples with a short-range
potential. The MAE method results in a nonlinear diffusion model
(which is thus local, with banded discretization matrix) for $p$, making it
straightforward to solve.

Recently \cite{Middleton:2014fa} argued in favour of the KSA because it gave the
two-particle density  as well as the one-particle density, and could
therefore capture correlations in particle positions, which the MFA
cannot. Our MAE method  
also gives an approximation for $P_2$, and it 
successfully captures the low likelihood of finding particles close to
each other when there are strong short-range repulsions. We emphasize
again that it does so at a fraction of the cost of KSA, so that the
MAE method becomes particularly suited for problems in 
two or three dimensions where the KSA or similar higher-order closures
are impractical.  We noted also that the MFA can be extended simply
(to what we called MFAe) to
produce an approximation for $P_2$, so that it too becomes a viable
alternative if the interactions are longer range.

Stochastic simulations of system of repulsive soft spheres are
generally regarded as simpler than the simulation of hard spheres
since one avoids the issue of how to approximate the collision between
two particles. Instead, soft sphere simulations involve summing up the
contribution to the interaction force of all the neighbors at each
time-step and adding it as a drift term to the Brownian
motion. However, for very repulsive potentials this needs to be done 
very carefully, since if the time-step is not small enough the
repulsive part of the potential is not resolved correctly and easily
missed. If this happened, the low-density diagonal in the two-particle
density plots (see for example \cref{fig:2dcase2}) would either not be
well resolved or not be there at all. It is therefore important to do
a proper convergence study for the stochastic simulations in order to
decide on the appropriate time-step. We note in this respect that our
coefficient $\alpha_u$ provides a natural way to determine the
radius of the equivalent hard-sphere  for any short range potential.

We have seen that the MAE method works well for repulsive short-range
potentials, while the MFA provides a good approximation for long-range
interactions. A natural question is to ask what to do for potentials
with both characteristics, namely those which are very singular at the
origin but with fat tails at infinity.   
This work provides a possible route to deal with such potentials: to
combine the MAE and MFA methods. Specifically, one could break the
potential into two 
parts and deal with each of them separately. The result would  be
an equation of the type \cref{mean_field} with the nonlocal
convolution term due to the long-range component of the potential, and
a nonlinear diffusion due to the short-range component.

\end{document}